\documentclass{aastex63}

\usepackage{amsmath}
\usepackage{times}
\usepackage{mathptmx}
\usepackage[caption=false]{subfig}
\DeclareMathAlphabet{\mathcal}{OMS}{cmsy}{m}{n}
\DeclareSymbolFont{largesymbols}{OMX}{cmex}{m}{n}

\received{\today}
\revised{\today}
\accepted{XXX}

\shorttitle{Correlations of GRBs} \shortauthors{Xu et al.}

\begin{document}

\title{Softness Ratio of SWIFT Gamma-ray Bursts and Relevant Correlations}

\correspondingauthor{Fan Xu (fanxu@ahnu.edu.cn), Yong-Feng Huang (hyf@nju.edu.cn)}

\author[0000-0001-7943-4685]{Fan Xu$^\star$}
\affiliation{Department of Physics, Anhui Normal University, Wuhu, Anhui 241002, People's Republic of China}
\affiliation{School of Astronomy and Space Science, Nanjing University,
    Nanjing 210023, People's Republic of China}

\author[0000-0001-7199-2906]{Yong-Feng Huang$^\star$}
\affiliation{School of Astronomy and Space Science, Nanjing University,
    Nanjing 210023, People's Republic of China}
\affiliation{Key Laboratory of Modern Astronomy and Astrophysics (Nanjing University),
    Ministry of Education, People's Republic of China}

\author{Liang Li}
\affiliation{Institute of Fundamental Physics and Quantum Technology, Ningbo University,
    Ningbo, Zhejiang 315211, People's Republic of China}
\affiliation{School of Physical Science and Technology, Ningbo University,
    Ningbo, Zhejiang 315211, People's Republic of China}
\affiliation{INAF-Osservatorio Astronomico d'Abruzzo, Via M.
Maggini snc, I-64100, Teramo, Italy}

\author{Jin-Jun Geng}
\affiliation{Purple Mountain Observatory, Chinese Academy of Sciences,
    Nanjing 210023, People's Republic of China}

\author{Xue-Feng Wu}
\affiliation{Purple Mountain Observatory, Chinese Academy of Sciences,
    Nanjing 210023, People's Republic of China}

\author{Song-Bo Zhang}
\affiliation{Purple Mountain Observatory, Chinese Academy of Sciences,
    Nanjing 210023, People's Republic of China}

\author{Chen Deng}
\affiliation{School of Astronomy and Space Science, Nanjing University,
    Nanjing 210023, People's Republic of China}

\author{Chen-Ran Hu}
\affiliation{School of Astronomy and Space Science, Nanjing University,
    Nanjing 210023, People's Republic of China}

\author{Xiao-Fei Dong}
\affiliation{School of Astronomy and Space Science, Nanjing University,
    Nanjing 210023, People's Republic of China}

\author{Hao-Xuan Gao}
\affiliation{Purple Mountain Observatory, Chinese Academy of Sciences,
    Nanjing 210023, People's Republic of China}

\begin{abstract}
The properties of X-ray flashes (XRFs) and X-ray rich gamma-ray
bursts (XRRs) as compared with classical gamma-ray bursts (C-GRBs)
have been widely discussed during the \emph{Swift} era. It has
been proposed that XRFs and XRRs are low-energy extensions of the
GRB population so that they should follow similar correlations. To
further examine this idea, we collect a sample of $303$ GRBs
detected by \emph{Swift} over the past two decades, all of which
have reliable redshifts and spectral parameters. The bursts are
classified into XRFs, XRRs, and C-GRBs based on their softness
ratio (SR), which is calculated by dividing the $25-50$ keV
fluence with the $50-100$ keV fluence.  A strong correlation is
found among the isotropic energy $E_{\mathrm{iso}}$, peak
luminosity $L_{\mathrm{p}}$, and rest frame burst duration $T_{90,
\mathrm{rest}}$, i.e., $E_{\mathrm{iso}} \propto
L_{\mathrm{p}}^{0.88\pm0.02} T_{90, \mathrm{rest}}^{0.58\pm0.02}$.
Additionally, two tight three-parameter correlations involving SR
and the rest-frame peak energy $E_{\mathrm{p}}$ are also derived,
i.e. $E_{\mathrm{p}} \propto E_{\mathrm{iso}}^{0.20\pm0.02}
\mathrm{SR}^{-2.27\pm0.15}$ and $E_{\mathrm{p}} \propto
L_{\mathrm{p}}^{0.17\pm0.02} \mathrm{SR}^{-2.33\pm0.14}$. It is
interesting to note that XRFs, XRRs, and C-GRBs all follow the
above correlations. The implications of these correlations and
their potential application in cosmology are discussed.
\end{abstract}


\section{Introduction} \label{sec:intro}

The prompt emission of gamma-ray bursts (GRBs) spans a broad
energy range from keV to GeV \citep{Abdo..2009,ZhangBB..2011}. It
is found that the softness ratios (SRs), defined as the fluence
ratio between the X-ray and gamma-ray bands, vary significantly
among bursts \citep{Barraud..2003,Sakamoto..2005}. Bursts with
softer spectra are classified as X-ray flashes (XRFs;
\cite{Heise..2001,Kippen..2001}), while those with harder spectra
are referred to as classical GRBs (C-GRBs;
\citep{Sakamoto..2008}). Between these two categories lies the
intermediate class of X-ray-rich GRBs (XRRs; \cite{Heise..2001}).
Generally speaking, XRFs and XRRs are regarded as low-energy
extensions of the GRB population \citep{Barraud..2005}.

The origin of XRFs and XRRs remains a topic of debate
\citep{Zhang..2002,Sakamoto..2008}. One possible explanation is
that intrinsic factors cause the differences between XRFs, XRRs,
and C-GRBs. It has been argued that XRFs may originate from GRB
jets with lower Lorentz factors, namely dirty fireballs
\citep{Dermer..1999,Huang..2002,Sun..2024,Geng..2025}. Alternatively, they might from
jets with low radiation efficiency
\citep{Mochkovitch..2004,Barraud..2005}. Additionally, it is
proposed that non-thermal emission from the dissipative process 
of the photosphere could account
for the strong X-ray emission in XRFs and XRRs
\citep{Ramirez-Ruiz..2005,Peer..2006}. On the other hand,
extrinsic factors have also been considered. Several groups have
investigated the effects of viewing angle or structured jets on
the observed softness of GRBs
\citep{Yamazaki..2002,Yamazaki..2004,Zhang..2004,Lamb..2004,Granot..2005,Xu..2023a,OConnor..2024,Gao..2024}.
Interestingly, these models are also commonly invoked to explain
orphan afterglows
\citep{Rhoads..1997,Huang..2002,Urata..2015,Xu..2023b,Ye..2024,Li..2024}.

The parameter distributions and correlations of XRFs, XRRs, and
C-GRBs have also been investigated in depth by various
researchers. No systematic differences are found for these three
subgroups in their low-energy photon indices, high-energy photon
indices, or durations
\citep{Kippen..2003,Sakamoto..2005,Sakamoto..2008}. Additionally,
all the three categories appear to follow similar empirical
correlations. For example, an anti-correlation has been found
between SR and the peak energy of the $\nu F_\nu$ spectra,
$E_{\mathrm{p, obs}}$
\citep{Sakamoto..2005,Sakamoto..2008,Katsukura..2020}. It
indicates that XRFs tend to have lower $E_{\mathrm{p, obs}}$.
Furthermore, the rest-frame peak energy $E_{\mathrm{p}}$ and the
isotropic energy $E_{\mathrm{iso}}$ exhibit a robust relationship,
known as the $E_{\mathrm{p}}-E_{\mathrm{iso}}$ (Amati) correlation
\citep{Amati..2002}. This correlation has been shown to hold
consistently across XRFs, XRRs, and C-GRBs
\citep{Lamb..2004b,Sakamoto..2006,Amati..2006,Bi..2018,Katsukura..2020}.
Several other correlations have also been explored, though their
robustness is often limited by the sample size and the uncertainty
of the data.

In this study, we conduct a detailed analysis of GRB prompt
emission correlations, paying special attention to GRBs that have
well-constrained spectra and redshift measurements. The GRB data
in our sample are taken from the third \emph{Swift} Burst Alert
Telescope (BAT; \cite{Barthelmy..2005a}) GRB
catalog\footnote{\url{https://swift.gsfc.nasa.gov/results/batgrbcat/index\_tables.html}}
(hereafter the BAT3 catalog ; \cite{Lien..2016}), including $303$
GRBs observed between $2005$ January and $2024$ September. The
\emph{Swift} mission \citep{Gehrels..2004}, launched in $2004$
with an initial two-year operational plan, has far exceeded
expectations. Over the past two decades, it has provided
groundbreaking GRB observations and revolutionized our
understanding of GRBs (e.g.,
\cite{Barthelmy..2005b,Gehrels..2005,Villasenor..2005,Zhang..2006,Gehrels..2006,Greiner..2009,Cucchiara..2011,Jin..2016}).
Using this rich data set, we will explore potential correlations
among various parameter pairs or combinations, which may offer new
insights into GRB physics and could be applied in various fields
such as cosmology.

The structure of this paper is as follows. In Section
\ref{sec:sample}, we describe our sample selection criteria and
data reduction procedures. Section \ref{sec:res} presents the
statistical analysis and results. In Section \ref{sec:concl}, we
summarize our conclusions and discuss their implications.
Throughout this study, a flat $\Lambda$CDM model with $H_{0}=67.8$
km s$^{-1}$ Mpc$^{-1}$ and $\Omega_{\mathrm{m}}=0.308$ is assumed
\citep{Planck..2016}.

\section{Sample selection and data analyze} \label{sec:sample}

The BAT3 catalog contains a total of $1585$ GRBs detected by
\emph{Swift} over the past two decades \citep{Lien..2016}. From
this catalog, we built our sample based on three main criteria.
($1$) Some key observational parameters, including the GRB
duration $T_{90}$, fluence, and peak flux, should be available.
This ensures that we can perform reliable correlation analyses.
($2$) The redshift should have been measured for each burst. ($3$)
The photon index and peak energy ($E_{\mathrm{p, obs}}$) should be
well-constrained for the event, as $E_{\mathrm{p, obs}}$ is a key
parameter in GRB correlation studies. Additionally, accurate
spectral information is necessary for performing k-corrections.
While the relatively narrow energy range of \emph{Swift} leads to
difficulty in measuring $E_{\mathrm{p, obs}}$ accurately, we
referred to the spectral parameters available from several other
sources, which include the \emph{Fermi} online
catalog\footnote{\url{https://heasarc.gsfc.nasa.gov/W3Browse/fermi/fermigbrst.html}}
\citep{Gruber..2014,Kienlin..2014,Bhat..2016,Kienlin..2020}, the
Konus-Wind online
catalog\footnote{\url{https://www.ioffe.ru/LEA/catalogs.html}}
\citep{Tsvetkova..2017,Tsvetkova..2021}, and the General
Coordinate Network (GCN)
Circulars\footnote{\url{https://gcn.gsfc.nasa.gov/circulars}}.
$15$ GRBs with well-constrained cut-off power-law spectra from the
BAT3 catalog are also included in our sample. After applying these
criteria, we finally built our sample that includes $303$ GRBs.
The observational parameters of these bursts taken from the Swift
database are presented in Table \ref{tab:1}, while the spectral
parameters of each burst are listed in Table \ref{tab:2}.

In Table \ref{tab:1}, we categorize our sample into long GRBs
(LGRBs) with $T_{90} > 2$ s and short GRBs (SGRBs) with $T_{90} <
2$ s. Note that six well-known GRBs with extended emission (i.e.
GRBs $061006$, $071227$, $070714$B, $090510$, $100816$A, and
$160410$A), which have $T_{90} > 2$ s, are classified as SGRBs
instead of LGRBs
\citep{Putten..2014,Lien..2016,Minaev..2020,LiXJ..2021}. This
results in $21$ SGRBs and $282$ LGRBs included in our sample. This
classification of long and short events proves important in
subsequent correlation analysis, as SGRBs exhibit distinct
behaviors compared to LGRBs in some correlations.

We classify XRFs, XRRs, and C-GRBs based on the following criteria proposed by \cite{Sakamoto..2008}

\begin{equation} \label{eq:1}
    \begin{aligned}
        S(25-50~\mathrm{keV}) / S(50-100~\mathrm{keV}) & \leq 0.72 & \quad(\mathrm{C}-\mathrm{GRB}), \\
        0.72 < S(25-50~\mathrm{keV}) / S(50-100~\mathrm{keV}) & \leq 1.32 & \quad(\mathrm{XRR}), \\
        1.32 < S(25-50~\mathrm{keV}) / S(50-100~\mathrm{keV}) & & \quad(\mathrm{XRF}) .
    \end{aligned}
\end{equation}
We define the softness ratio as $\mathrm{SR} =
S(25-50~\mathrm{keV}) / S(50-100~\mathrm{keV})$, which is the
ratio of fluence in the $25-50$ keV band ($S(25-50~\mathrm{keV})$)
with respect to that in $50-100$ keV ($S(50-100~\mathrm{keV})$).
The values of $S(25-50~\mathrm{keV})$ and $S(50-100~\mathrm{keV})$
are taken from the BAT3 catalog \citep{Lien..2016}. Among the
$303$ GRBs in our sample, we have $5$ XRFs, $167$ XRRs, and $131$
C-GRBs according to the above definition. The distribution of SR
for the entire sample is shown in Figure \ref{fig:2}. The number
of XRFs in our sample is quite limited due to our requirement for
a well-constrained spectrum. XRFs generally have $E_{\mathrm{p,
obs}} < 30$ keV, which is very close to the low-energy threshold
of most GRB detectors (e.g., \emph{Fermi}/GBM, Konus-\emph{Wind},
and \emph{Swift}/BAT). Therefore the spectral parameters of XRFs
are often poorly constrained. In contrast, the proportion of XRRs
(55\%) and C-GRBs (43\%) in our sample are comparable, which is
consistent with previous studies
\citep{Sakamoto..2008,Bi..2018,Katsukura..2020}.

With the energy fluence listed in Table \ref{tab:1}, we derive the
isotropic energy of the prompt emission as

\begin{equation} \label{eq:2}
    E_{\mathrm{iso}}=\frac{4 \pi D_{\mathrm{L}}^{2}(z) S(15-150~\mathrm{keV}) k}{1+z},
\end{equation}
where $S(15-150~\mathrm{keV})$ is the energy fluence in $15-150$
keV band, $z$ is the redshift and $D_{\mathrm{L}}$ is the
luminosity distance. $k$ represents the k-correction factor, which
adjusts the energy to a common cosmological rest-frame band of
$1-10^{4}$ keV. To calculate the k-correction factor, we use the
spectra information listed in Table \ref{tab:2}. For each burst in
our sample with a photon spectrum of $N(E)$, the k-correction
factor $k$ is given by

\begin{equation}
    k = \frac{\int_{1 /(1+z)}^{10^{4} /(1+z)} E N(E) d E}{\int_{15}^{150} E N(E) d E}.
\end{equation}
We can also calculate the isotropic peak luminosity with $k$, i.e.,

\begin{equation}
    L_{\mathrm{p}}=4 \pi D_{\mathrm{L}}^{2} F(15-150~\mathrm{keV}) k,
\end{equation}
where $F(15-150~\mathrm{keV})$ is the peak energy flux in $15-150$ keV band.

The parameters relevant to our correlation study are summarized in
Table \ref{tab:3}. These include the softness ratio SR, the rest
frame burst duration $T_{90, \mathrm{rest}}=T_{90}/(1+z)$, the
isotropic energy $E_{\mathrm{iso}}$, the isotropic peak luminosity
$L_{\mathrm{p}}$, and the rest frame peak energy
$E_{\mathrm{p}}=E_{\mathrm{p, obs}}\times(1+z)$. With these data,
we explore the parameter distributions and correlations for XRFs,
XRRs, and C-GRBs. The results of these analyses are presented in
the following section.

\section{Results} \label{sec:res}

\subsection{Distributions of Parameters}
\label{sec:test}

\begin{figure}[htbp]
    \centering
    \includegraphics[width=0.8\linewidth]{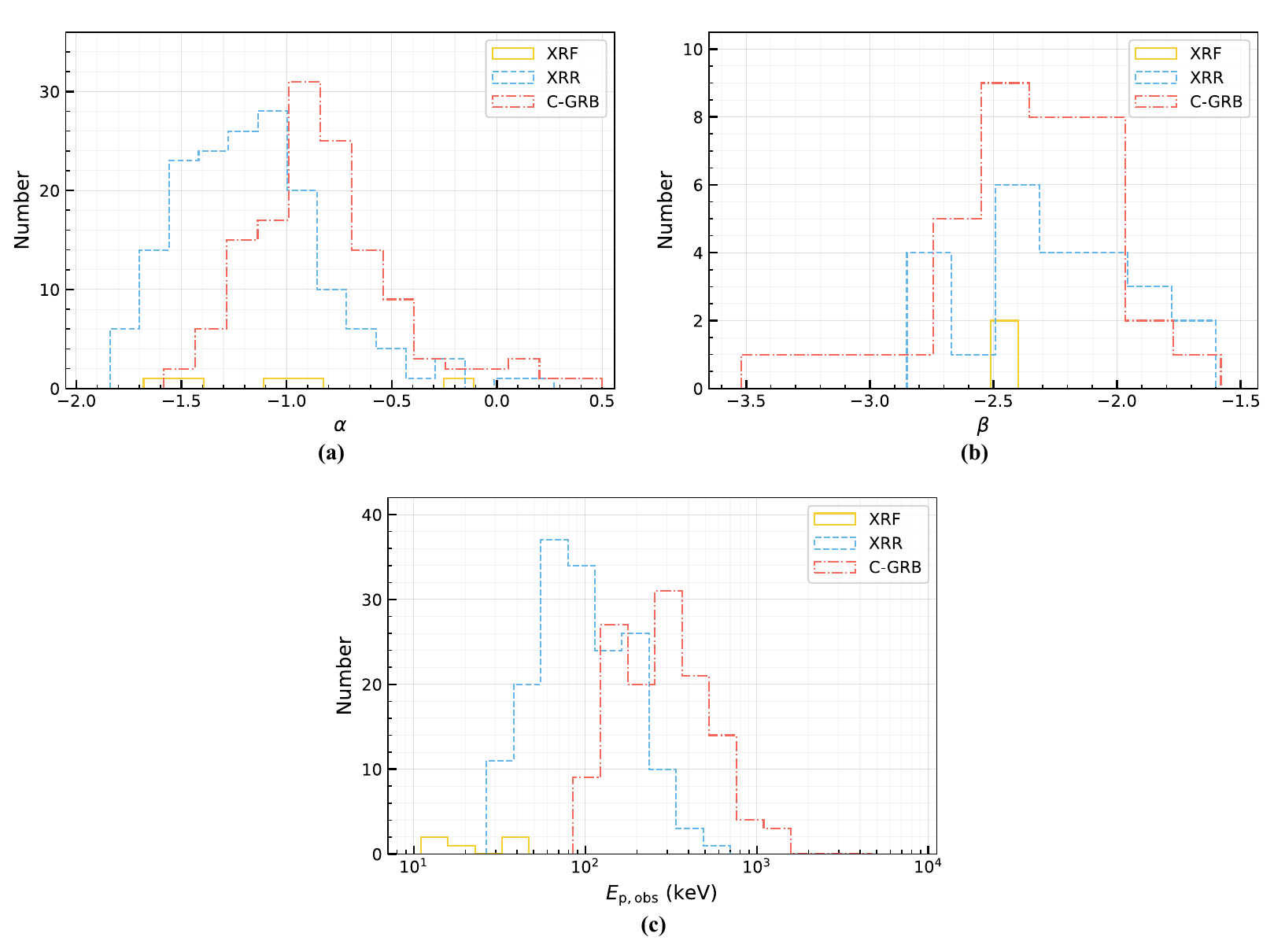}
    \caption{The distribution of $\alpha$ (Panel (a)), $\beta$ (Panel (b)),
    and $E_{\mathrm{p, obs}}$ (Panel (c)) for the whole sample.
    XRFs, XRRs, and C-GRBs are marked with yellow solid lines, blue dashed lines,
    and red dash-dotted lines, respectively. } \label{fig:2}
\end{figure}

In Figure \ref{fig:2}, we present the distributions of spectral
parameters for XRFs, XRRs, and C-GRBs. As shown in Figure
\ref{fig:2}(c), the observed peak energy varies significantly
among XRFs, XRRs, and C-GRBs. This is consistent with previous
studies
\citep{Sakamoto..2005,Sakamoto..2008,Bi..2018,Katsukura..2020}.
The low-energy index $\alpha$ in our sample ranges from $-1.84$ to
$0.5$ and peaks around $-1$. The distribution of the high energy
index $\beta$ in our sample clusters around $-2.5$ and ranges in
($-3.52, -1.58$). Notably, as shown in Figure \ref{fig:2}(a), the
$\alpha$ values of XRRs are systematically lower than those of
C-GRBs. However, no significant difference is observed in the
high-energy index. This contrasts with earlier studies
\citep{Kippen..2003,Sakamoto..2005}, where no systematic
differences were found for both $\alpha$ and $\beta$. The lower
values of $\alpha$ for XRRs are consistent with their
classification as being softer than C-GRBs in the low-energy band.

To further investigate the difference of $\alpha$ among XRFs,
XRRs, and C-GRBs, we have performed pairwise Kolmogorov-Smirnov
(K-S) tests. The results are presented in Table \ref{tab:4}. For
XRRs and C-GRBs, we find the K-S test probability ($P$) of
$\alpha$ is $3.70 \times 10^{-13}$, indicating a significant
difference in the low-energy indices between the two categories.
In contrast, the K-S test probability of $\beta$ is $0.8$,
suggesting that the high-energy indices of XRRs and C-GRBs are
statistically indistinguishable. Since the sample size of XRFs is
too small, K-S tests involving XRFs fail to generate any
statistically reliable conclusions.

K-S tests are also performed to examine the statistical difference
between LGRBs and SGRBs. It is found that both the observed peak
energy ($E_{\mathrm{p, obs}}$) and the low-energy index ($\alpha$)
differ significantly between LGRBs and SGRBs, with $P \sim
0.00032$ and $P \sim 3.10 \times 10^{-05}$, respectively.
Specifically, SGRBs tend to exhibit higher $E_{\mathrm{p, obs}}$
and $\alpha$, suggesting that they are harder than LGRBs. However,
since none of the SGRBs in our sample have well-constrained
$\beta$ values, we cannot assess whether the high-energy index
differs between LGRBs and SGRBs.

In Table \ref{tab:4}, we also present the $P$-values for key
parameters relevant to our correlation study. For instance, the
rest-frame peak energy $E_{\mathrm{p}}$ is similar for LGRBs and
SGRBs ($P=0.26$). This could result from a selection effect, as
SGRBs usually have lower redshifts and higher observed peak
energies compared to LGRBs. In contrast, both the isotropic energy
($E_{\mathrm{iso}}$) and the peak luminosity ($L_{\mathrm{p}}$) of
LGRBs are systematically higher than that of SGRBs. Furthermore,
C-GRBs exhibit larger $E_{\mathrm{iso}}$ and $L_{\mathrm{p}}$ than
XRRs. However, the rest frame durations of XRRs and C-GRBs are
very similar and indistinguishable.

\subsection{Correlations among parameters}
\label{sec:relation}

We have explored potential correlations between various parameter
combinations. To get the best-fit expressions for the
correlations, the Markov chain Monte Carlo (MCMC) algorithm is
used. The Spearman correlation coefficient $\rho$ and the
corresponding chance probability $p$ are calculated for the
best-fit curve. The results are summarized in Table \ref{tab:5}.

\subsubsection{The Correlations between $E_{\mathrm{iso}}$, $L_{\mathrm{p}}$, and $T_{90, \mathrm{rest}}$} \label{sec:LTE-relation}

\begin{figure}[htbp]
    \centering
    \includegraphics[width=0.8\linewidth]{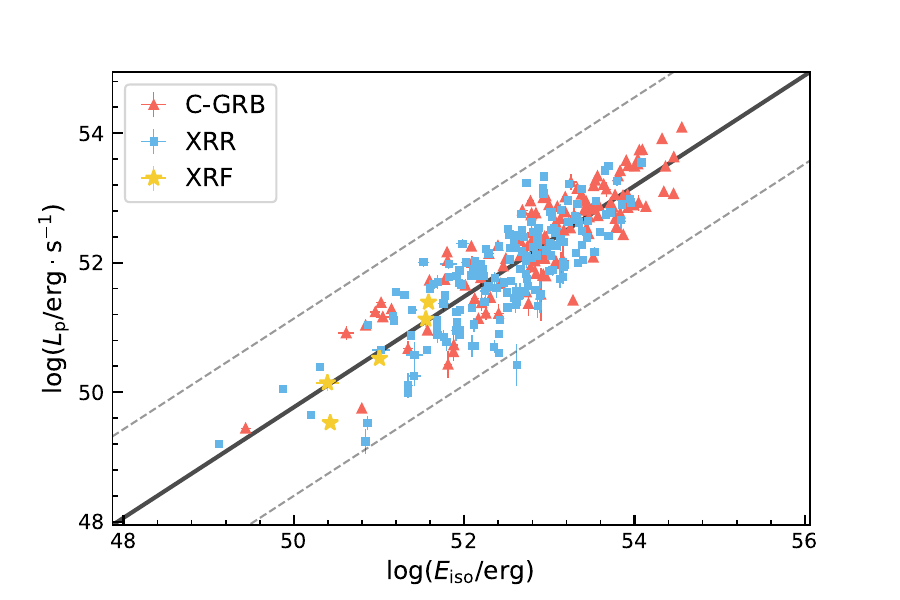}
    \caption{The best-fit $E_{\mathrm{iso}}-L_{\mathrm{p}}$ correlation for the $303$ GRBs in the
    sample. XRFs, XRRs, and C-GRBs are marked with yellow stars, blue squares, and red triangles,
    respectively. The black solid line is the best-fit curve for the observational data points.
    The gray dashed lines represent the $3\sigma$ confidence level. }
    \label{fig:3}
\end{figure}

Previous studies have revealed a potential correlation between
$E_{\mathrm{iso}}$ and $L_{\mathrm{p}}$
\citep{Lamb..2005,Ghirlanda..2005}. In Figure \ref{fig:3}, we plot
$E_{\mathrm{iso}}$ versus $L_{\mathrm{p}}$ for all GRBs in our
samples. A clear positive correlation is observed, indicating that
more energetic burst tend to have a higher peak luminosity, which
aligns with expectations. The best-fit correlation is
$L_{\mathrm{p}} \propto E_{\mathrm{iso}}^{0.85\pm0.03}$, with
$\rho = 0.85$ and $p = 7.60 \times 10^{-84}$.

\begin{figure}[htbp]
    \centering
    \includegraphics[width=0.8\linewidth]{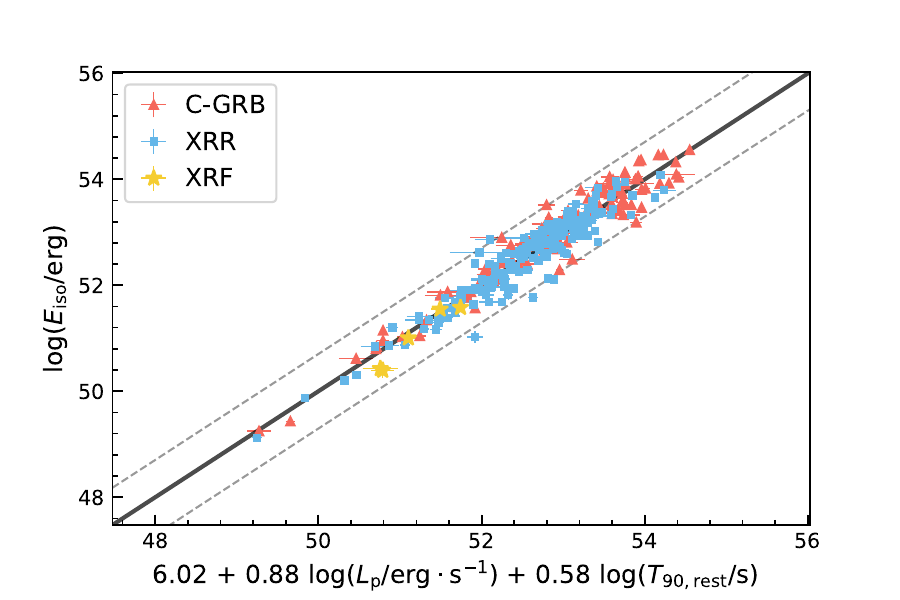}
    \caption{The best-fit $L_{\mathrm{p}}-T_{90, \mathrm{rest}}-E_{\mathrm{iso}}$ correlation
    for the $303$ GRBs in the sample. XRFs, XRRs, and C-GRBs are marked with yellow stars,
    blue squares, and red triangles, respectively. The black solid line is the best-fit curve
    for the observational data points. The gray dashed lines mark the $3\sigma$ confidence range. }
    \label{fig:4}
\end{figure}

However, the intrinsic scatter of this correlation is relatively
large ($\sigma_{\mathrm{in}}=0.46\pm0.02$). This may be due to the
fact that the total energy released in a burst is not solely
determined by the peak luminosity but also influenced by the burst
duration, i.e., $E_{\mathrm{iso}} \sim L_{\mathrm{p}} \times
T_{90, \mathrm{rest}}$. Inspired by this equation, we have
explored the three-parameter correlation involving
$E_{\mathrm{iso}}$, $L_{\mathrm{p}}$, and $T_{90, \mathrm{rest}}$.
The best-fit result is $E_{\mathrm{iso}} \propto
L_{\mathrm{p}}^{0.88\pm0.02} T_{90, \mathrm{rest}}^{0.58\pm0.02}$
with a much smaller intrinsic scatter of
$\sigma_{\mathrm{in}}=0.24\pm0.01$ as compared with the
two-parameter $E_{\mathrm{iso}}-L_{\mathrm{p}}$ correlation. The
Spearman correlation coefficient for this three-parameter relation
is $\rho=0.95$, with a probability of $p = 6.20 \times 10^{-153}$,
suggesting a strong correlation among $E_{\mathrm{iso}}$,
$L_{\mathrm{p}}$, and $T_{90, \mathrm{rest}}$. Our fitting result
is illustrated in Figure \ref{fig:4}.

From Figure \ref{fig:4}, we see that both LGRBs and SGRBs, as well
as XRFs, XRRs, and C-GRBs, are tightly distributed along the
best-fit line. It is striking to see that this correlation holds
even with a large sample size of $303$ GRBs, and it applies for
all these different subclasses of events. The power-law index of
$L_{\mathrm{p}}$ in the correlation is close to $1$, which is
consistent with the expectation from $E_{\mathrm{iso}} \sim
L_{\mathrm{p}} \times T_{90, \mathrm{rest}}$. However, the index
of $T_{90, \mathrm{rest}}$ is around $0.6$, which is much less
than $1$. This deviation can be attributed to the multiple spike
structure often observed in the GRB light curves of prompt
emission. Consequently, the isotropic energy integrated over the
burst duration ($E_{\mathrm{iso}}$) is typically smaller than
$L_{\mathrm{p}} \times T_{90, \mathrm{rest}}$.

Interestingly, \cite{Xu..2012} found a tight \emph{L-T-E}
correlation for GRBs with a plateau in the X-ray afterglow light
curve. It is expressed as $L_{X} \propto
T_{a}^{-0.99}E_{\mathrm{iso}}^{0.86}$
\citep{Tang..2019,Zhao..2019,Deng..2023}, where $T_{a}$ is the
plateau duration and $L_{X}$ is the X-ray luminosity of the
plateau. Such an \emph{L-T-E} correlation involving
$L_{\mathrm{p}}$, $T_{90, \mathrm{rest}}$ and $E_{\mathrm{iso}}$
is further confirmed in our current study.

\begin{figure}[htbp]
    \centering
    \includegraphics[width=0.8\linewidth]{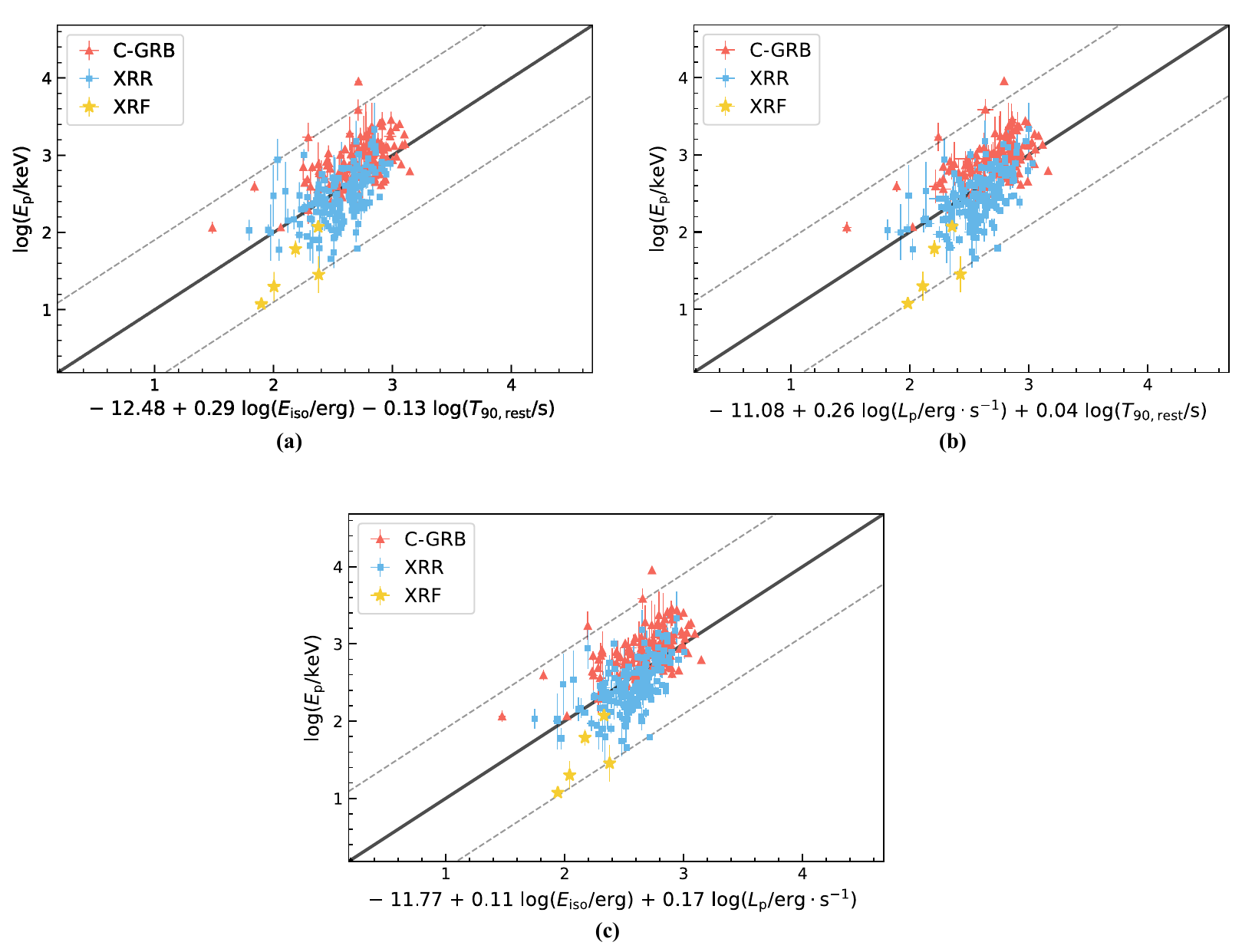}
    \caption{The best-fit results of the $E_{\mathrm{iso}}-T_{90, \mathrm{rest}}-E_{\mathrm{p}}$
    correlation (Panel (a)), $L_{\mathrm{p}}-T_{90, \mathrm{rest}}-E_{\mathrm{p}}$
    correlation (Panel (b)), and $E_{\mathrm{iso}}-L_{\mathrm{p}}-E_{\mathrm{p}}$
    correlation (Panel (c)). XRFs, XRRs, and C-GRBs are marked with yellow stars,
    blue squares, and red triangles, respectively. The black solid lines represent the best-fit
    curves of the correlations. The gray dashed lines mark the $3\sigma$ confidence range. }
    \label{fig:5}
\end{figure}

Additionally, we have also explored other three parameter
correlations such as $E_{\mathrm{iso}}-T_{90,
\mathrm{rest}}-E_{\mathrm{p}}$, $L_{\mathrm{p}}-T_{90,
\mathrm{rest}}-E_{\mathrm{p}}$, and
$E_{\mathrm{iso}}-L_{\mathrm{p}}-E_{\mathrm{p}}$. The best-fit
results are shown in Figure \ref{fig:5}. The intrinsic scatters of
these correlations are around $0.3$, with the Spearman correlation
coefficients of approximately $0.6$. It indicates that while these
parameters are truly correlated, the significance of these
correlations is generally not high. These correlations are likely
connected with the well-known Amati correlation
($E_{\mathrm{iso}}-E_{\mathrm{p}}$) \citep{Amati..2002} and the
Yonetoku correlation ($L_{\mathrm{p}}-E_{\mathrm{p}}$)
\citep{Yonetoku..2004}.

\subsubsection{The $E_{\mathrm{iso}}$-$\mathrm{SR}$-$E_{\mathrm{p}}$ and $L_{\mathrm{p}}$-$\mathrm{SR}$-$E_{\mathrm{p}}$ Correlations} \label{sec:SR-relation}

\begin{figure}[htbp]
    \centering
    \includegraphics[width=0.48\linewidth]{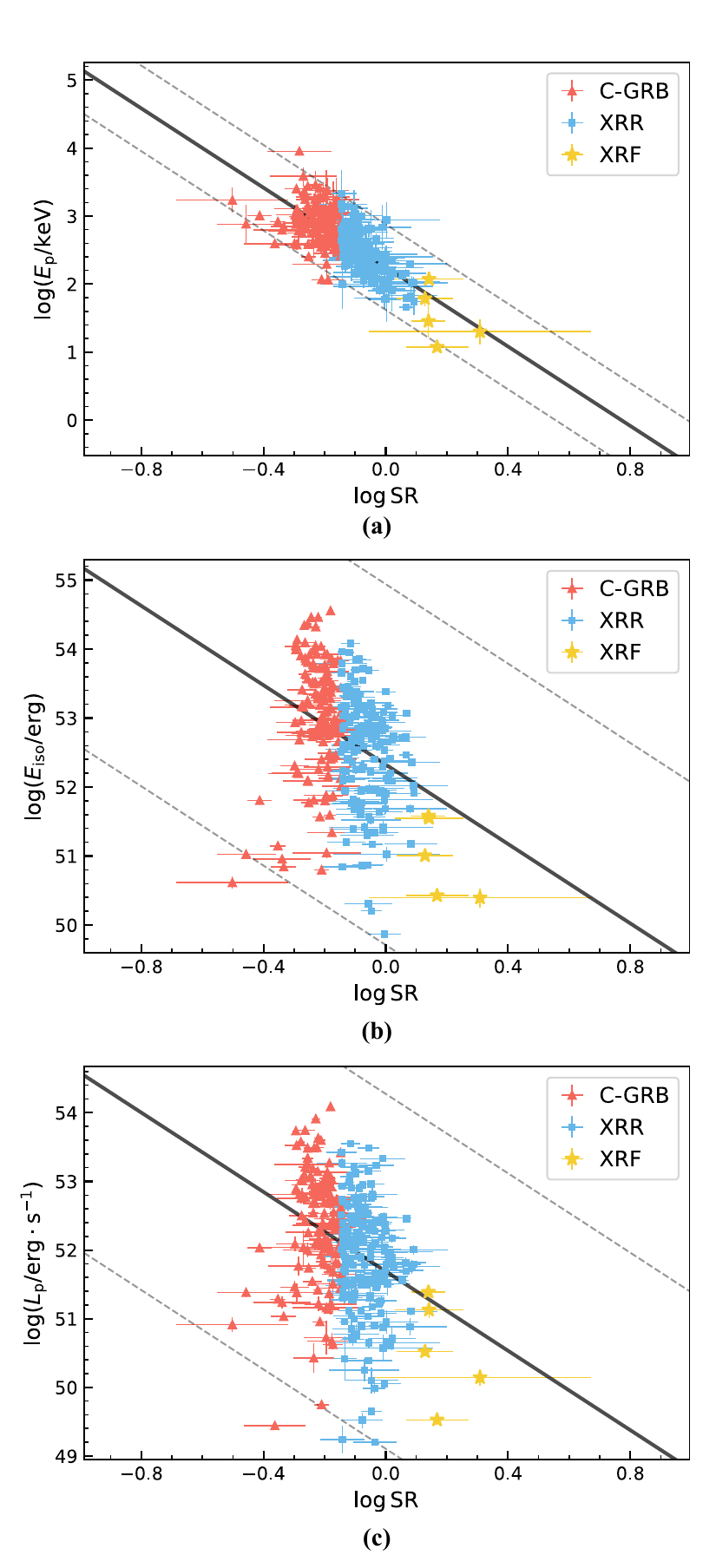}
    \caption{Dependence of $E_{\mathrm{p}}$, $E_{\mathrm{iso}}$, and $L_{\mathrm{p}}$ on SR.
    Panel (a) shows $E_{\mathrm{p}}$ vs. SR. Panel (b) shows $E_{\mathrm{iso}}$ vs. SR.
    Panel (c) shows $L_{\mathrm{p}}$ vs. SR. XRFs are marked with yellow stars,
    XRRs with blue squares, and C-GRBs with red triangles. The black solid lines illustrate
    the best-fit curves of the correlations, and the gray dashed lines represent
    the $3\sigma$ confidence interval. }
    \label{fig:6}
\end{figure}

It has been proposed that a strong correlation should exist
between the softness ratio and the peak energy
\citep{Sakamoto..2005,Sakamoto..2008,Katsukura..2020}.
\cite{Sakamoto..2008} calculated the theoretical SR curve as a
function of $E_{\mathrm{p, obs}}$. They used the Band function to
describe the $\gamma$-ray spectrum, taking the power-law indices
as $\alpha=-1$ and $\beta=-2.5$. It was found that the
observational data points distributed tightly around this curve.
The scatter of this correlation is primarily attributed to
variations of $\alpha$ and $\beta$ for individual bursts. In
Figure \ref{fig:6}(a), we plot $E_{\mathrm{p}}$ versus SR for the
GRBs in our sample. We can also see a strong inverse relationship
between the two parameters. The best-fit result is $E_{\mathrm{p}}
\propto \mathrm{SR}^{-2.91\pm0.16}$, with an intrinsic scatter of
$\sigma_{\mathrm{in}}=0.21\pm0.01$. The Spearman correlation
coefficient is $\rho=-0.74$, with a probability of $p = 1.30
\times 10^{-53}$. Such an anti-correlation is easy to understand
since a higher SR value corresponds to a softer spectrum and a
lower peak energy. In Figure \ref{fig:6}, we also show the
$\mathrm{SR}-E_{\mathrm{iso}}$ and $\mathrm{SR}-L_{\mathrm{p}}$
correlations, both of which are relatively diffuse.

\begin{figure}[htbp]
    \centering
    \includegraphics[width=0.8\linewidth]{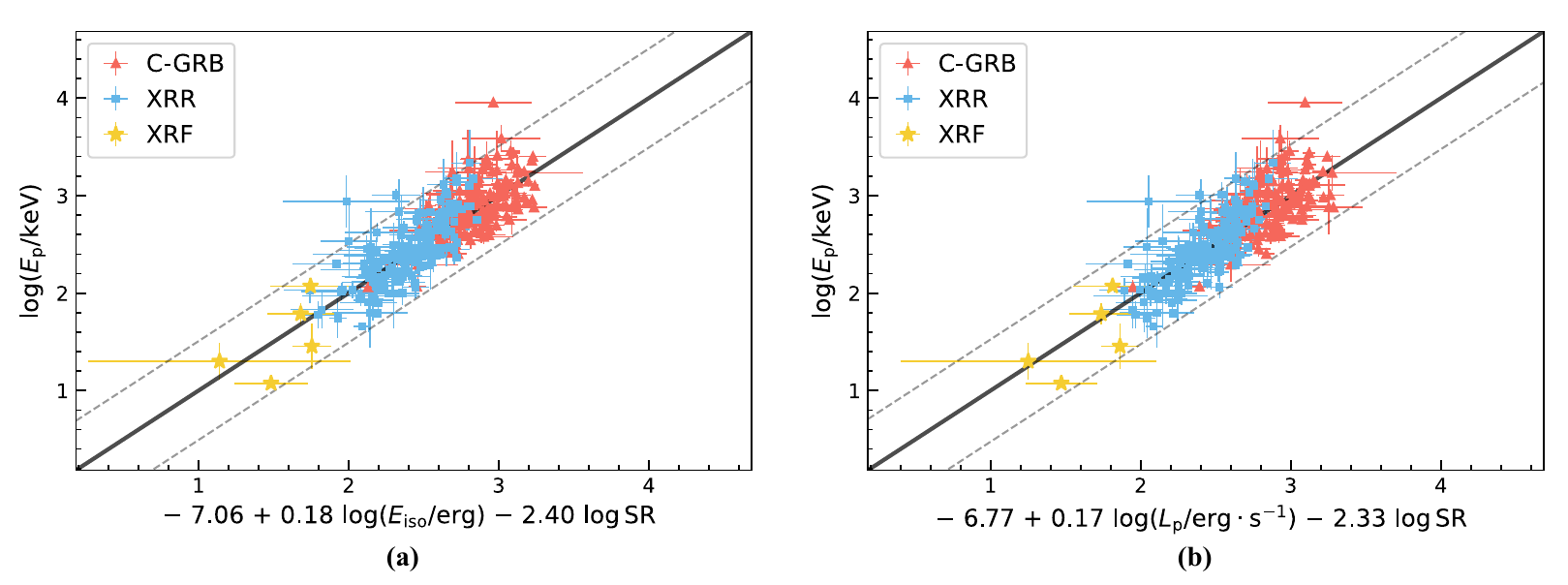}
    \caption{The best-fit $E_{\mathrm{iso}}-\mathrm{SR}-E_{\mathrm{p}}$ (Panel (a))
     and $L_{\mathrm{p}}-\mathrm{SR}-E_{\mathrm{p}}$ (Panel (b)) correlations for
     all of the $303$ GRBs in the sample. XRFs, XRRs, and C-GRBs are marked with
     yellow stars, blue squares, and red triangles, respectively. The black solid
     lines show the best-fit curves of the correlations, and the gray dashed lines
     mark the $3\sigma$ confidence interval. }
     \label{fig:7}
\end{figure}

On the other hand, the three-parameter correlations of
$E_{\mathrm{iso}}-\mathrm{SR}-E_{\mathrm{p}}$ and
$L_{\mathrm{p}}-\mathrm{SR}-E_{\mathrm{p}}$ are found to be very
tight. The best-fit results for these two three-parameter
correlations are shown in Figure \ref{fig:7}. Our best-fit
expressions are $E_{\mathrm{p}} \propto
E_{\mathrm{iso}}^{0.18\pm0.02} \mathrm{SR}^{-2.40\pm0.13}$, with
$\rho=0.84$ and $p = 2.70 \times 10^{-80}$, and $E_{\mathrm{p}}
\propto L_{\mathrm{p}}^{0.17\pm0.02} \mathrm{SR}^{-2.33\pm0.14}$,
with $\rho=0.84$ and $p=1.50 \times 10^{-80}$.

\begin{figure}[htbp]
    \centering
    \includegraphics[width=0.8\linewidth]{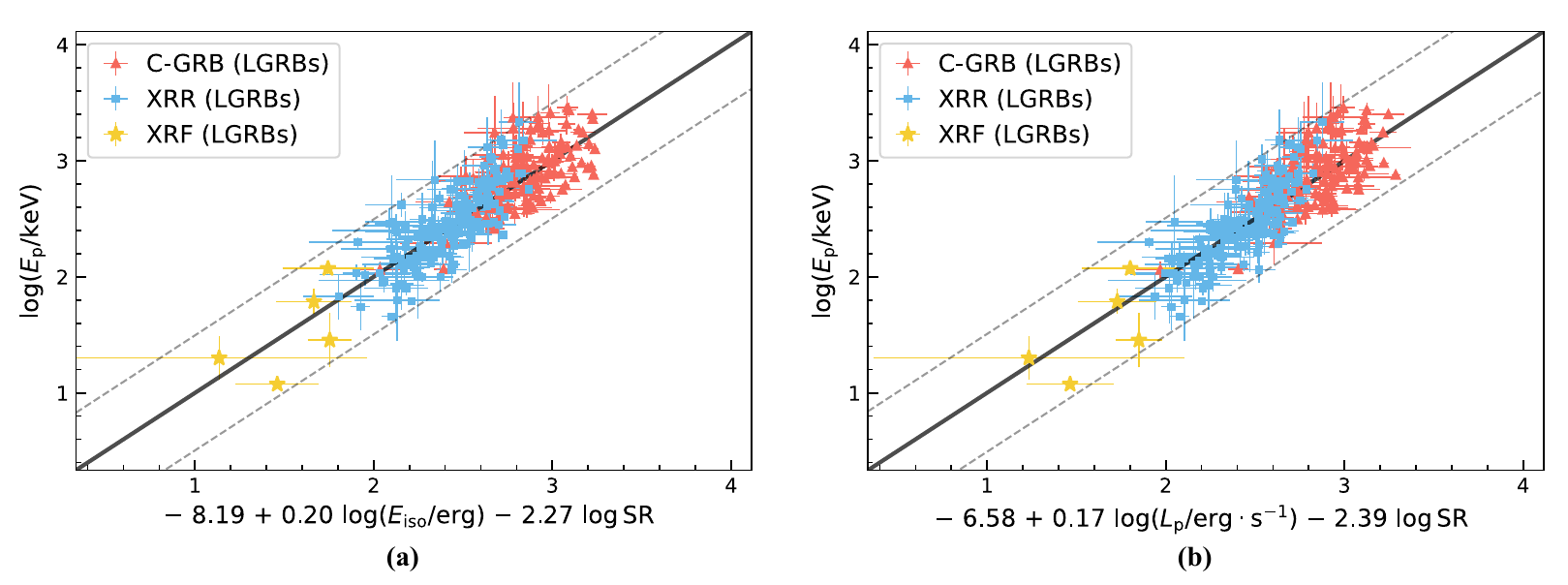}
    \caption{The best-fit $E_{\mathrm{iso}}-\mathrm{SR}-E_{\mathrm{p}}$ (Panel (a))
    and $L_{\mathrm{p}}-\mathrm{SR}-E_{\mathrm{p}}$ (Panel (b)) correlations
    for $282$ LGRBs in the sample. XRFs, XRRs, and C-GRBs are represented by
    yellow stars, blue squares, and red triangles, respectively. The black
    solid lines indicate the best-fit curves of the correlations, and
    the gray dashed lines denote the $3\sigma$ confidence interval.}
    \label{fig:8}
\end{figure}

Notably, SGRBs $090510$ and $071227$ appear as outliers in both
correlations. It indicates that LGRBs and SGRBs may follow
different $E_{\mathrm{iso}}-\mathrm{SR}-E_{\mathrm{p}}$ and
$L_{\mathrm{p}}-\mathrm{SR}-E_{\mathrm{p}}$ correlations. To be
more specific, we examine these two correlations for LGRBs in our
sample. For the $E_{\mathrm{iso}}-\mathrm{SR}-E_{\mathrm{p}}$
correlation, we get $E_{\mathrm{p}} \propto
E_{\mathrm{iso}}^{0.20\pm0.02} \mathrm{SR}^{-2.27\pm0.15}$, with
$\sigma_{\mathrm{in}}=0.16\pm0.01$. For the
$L_{\mathrm{p}}-\mathrm{SR}-E_{\mathrm{p}}$ correlation, we obtain
$E_{\mathrm{p}} \propto L_{\mathrm{p}}^{0.17\pm0.02}
\mathrm{SR}^{-2.33\pm0.14}$, with
$\sigma_{\mathrm{in}}=0.17\pm0.01$. The best-fit results are shown
in Figure \ref{fig:8}. Again, we see that XRFs, XRRs, and C-GRBs
follow the same tight correlation.

\begin{figure}[htbp]
    \centering
    \includegraphics[width=0.8\linewidth]{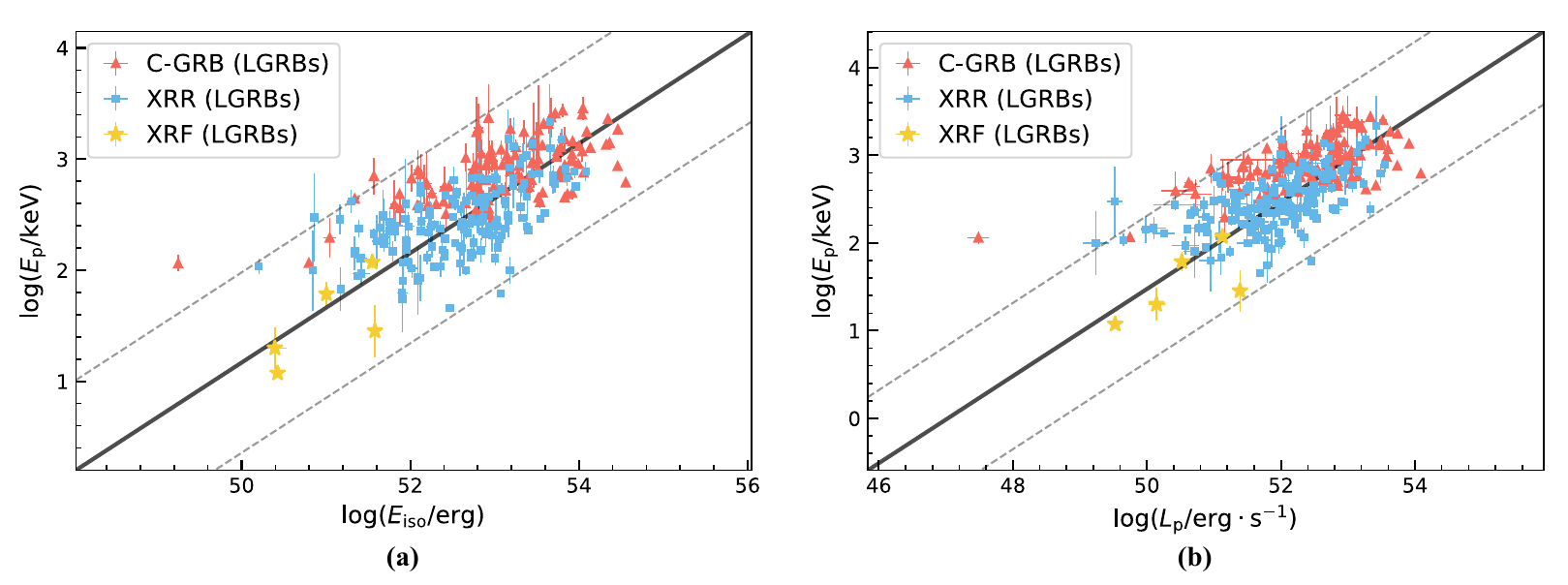}
    \caption{The best-fit results of $E_{\mathrm{iso}}-E_{\mathrm{p}}$
    and $L_{\mathrm{p}}-E_{\mathrm{p}}$ correlations for LGRBs in the
    sample. The black solid lines indicate the best-fit curves of the
    correlations, and the gray dashed lines denote the $3\sigma$ confidence range. }
    \label{fig:9}
\end{figure}

We have also examined the Amati correlation and the Yonetoku
correlation for LGRBs. The best-fit results are illustrated in
Figure \ref{fig:9}. We see that most GRBs follow the best-fit
lines represented by $E_{\mathrm{p}} \propto
E_{\mathrm{iso}}^{0.49\pm0.02}$ and $E_{\mathrm{p}} \propto
L_{\mathrm{p}}^{0.50\pm0.02}$. However, a few low-luminosity XRRs
and C-GRBs seem to follow a shallower track with significantly
smaller power-law indices. Currently, the origin of low-luminosity
GRBs is still under debate \citep{Dong..2023}. One possibility is
that they form a distinct GRB population \citep{Liang..2007}.
Another possibility is that they may be normal GRB outflows
observed off-axis
\citep{Granot..2002,Huang..2002,Yamazaki..2003,Xu..2023a,LiXJ..2024}.
Anyway, they exhibit some different properties compared to
high-luminosity GRBs, as shown in Figure \ref{fig:9}.

Comparing Figure \ref{fig:8} with Figure \ref{fig:9}, we find that
our three-parameter correlations are much tighter.  These strong
three-parameter correlations highlight the importance of SR in
linking $E_{\mathrm{p}}$ with both $E_{\mathrm{iso}}$ and
$L_{\mathrm{p}}$. Moreover, we have also explored whether
additional parameters could further tighten our correlations. For
example, we examined possible connection in the four parameters of
$E_{\mathrm{iso}}-T_{90,
\mathrm{rest}}-\mathrm{SR}-E_{\mathrm{p}}$, and in
$L_{\mathrm{p}}-T_{90, \mathrm{rest}}-\mathrm{SR}-E_{\mathrm{p}}$.
The best-fit results correspond to an intrinsic scatter of
$\sigma_{\mathrm{in}} \sim 0.16$ and $\rho=0.86$ for both
correlations and the power-law index of $T_{90, \mathrm{rest}}$ is
very close to $0$. It means that the burst duration plays a minor
role in these correlations so that these four-parameter
correlations are not preferred.

\subsubsection{Redshift evolution}

The parameters discussed above may be affected by selection biases due to instrumental thresholds and redshift evolution \citep{Dainotti..2013,Dainotti..2015}. The redshift evolution may distort the statistical correlations \citep{Dainotti..2022qusar}. To mitigate these biases in truncated datasets, the method proposed by \cite{Efron..1992} is commonly used (see also \cite{Dainotti..2017A&A,Dainotti..2020ApJ,Xu..2021}). It applies a modified version of Kendall's $\tau$ statistics and can effectively remove the redshift evolution from the parameters \citep{Dainotti..2021galaxy}. 

For a selected parameter $x$, we take the redshift evolution form as $g(z)=(1+z)^{k_x}$, where $k_x$ is the evolutionary coefficient. The de-evolved parameter is then given by $x^{\prime}=x/g(z)$ \citep{Dainotti..2023,Lenart..2025}. For the SR parameter, we expect no redshift evolution, as there is no redshift-related term in its equation. For other parameters, we use the results summarized in Table 2 by \cite{Dainotti..2022} as proxies. After de-evolving the parameters, we re-fit the correlations and summarize the results in Table \ref{tab:6}. 

Our analysis shows that most correlations exhibit a smaller intrinsic scatter after accounting for redshift effects. It indicates that these correlations become tighter when the redshift evolution is corrected. In contrast, a few correlations exhibit a slight increase in scatter following the correction. In particular, the $E_{\mathrm{iso}}-L_{\mathrm{p}}-T_{90, \mathrm{rest}}$ correlation shows a larger intrinsic scatter for both the entire sample and the LGRB sample. Overall, we note that for correlations with a relatively small intrinsic scatter ($\sigma_{\mathrm{in}} \lesssim 0.3$), the redshift correction appears to have only a modest effect on their tightness.

\section{Conclusions and discussion} \label{sec:concl}

In this study, we collect $303$ GRBs detected by \emph{Swift} with
known redshift and well-constrained spectra parameters. A
systematical statistical analysis is carried out based on the
sample. The GRBs are categorized into subgroups of XRFs, XRRs, and
C-GRBs, as well as LGRBs and SGRBs, to investigate and compare
their difference in parameter distributions and correlations.
Based on the K-S tests, it is found that XRRs and C-GRBs exhibit
obviously different low-energy indices and peak energies. While
SGRBs typically have lower redshifts and higher observed peak
energies, their rest-frame peak energies are similar to that of
LGRBs. A strong correlation is found among the three parameters of
$L_{\mathrm{p}}$, $T_{90, \mathrm{rest}}$, and $E_{\mathrm{iso}}$.
The best-fit expression is $E_{\mathrm{iso}} \propto
L_{\mathrm{p}}^{0.88\pm0.02} T_{90, \mathrm{rest}}^{0.58\pm0.02}$,
with an intrinsic scatter of $\sigma_{\mathrm{in}}=0.24\pm0.01$.
This correlation holds for various subgroups, including LGRBs and
SGRBs, as well as XRFs, XRRs, and C-GRBs. Additionally, two very
tight correlations involving the softness ratio (SR) are
identified. For LGRBs, we obtain $E_{\mathrm{p}} \propto
E_{\mathrm{iso}}^{0.20\pm0.02} \mathrm{SR}^{-2.27\pm0.15}$ and
$E_{\mathrm{p}} \propto L_{\mathrm{p}}^{0.17\pm0.02}
\mathrm{SR}^{-2.33\pm0.14}$. These three-parameter correlations
could be regarded as an extension of the well-known Amati and
Yonetoku correlations, but are interestingly much tighter, with an
intrinsic scatter of $\sigma_{\mathrm{in}} \sim 0.16$.
%% Several correlations between GRB prompt emission parameters
%% are explored in this study.

It is found that XRFs, XRRs, and C-GRBs follow similar
correlations involving prompt emission parameters. It supports the
idea that XRFs and XRRs are low-energy extensions of the GRB
population. The $L_{\mathrm{p}}-T_{90,
\mathrm{rest}}-E_{\mathrm{iso}}$ correlation involves parameters
that are closely related to the temporal property of GRB prompt
emission. The existence of this correlation indicates that GRB
prompt emission light curves may share similar pulse profiles
despite their apparent complexity. For instance, a fast-rising
exponential-decay profile may dominate the light curves of most
GRBs \citep{Norris..1996}. Furthermore, since both SGRBs and LGRBs
adhere to this correlation, these similar structures may exist in
both SGRBs and LGRBs.

In our study, the softness ratio (SR) is defined as the ratio of
fluence in the $25-50$ keV band with respect to that in the
$50-100$ keV band. The logarithm value of SR employed in our
correlation analysis is somewhat similar to the color parameter
used in the standard candle correlation of Type Ia supernovae (SNe
Ia). In the Tripp formula for SNe Ia, a color term of $b$(B-V) is
included, where $b$ is the correlation coefficient and (B-V)
represents the color index \citep{Tripp..1998}. The inclusion of
this term significantly tightens the SNe Ia correlation.
Similarly, we find that both the Amati and Yonetoku correlations
become much tighter when the SR parameter is included.

While we discuss the correlation concerning the prompt emission phase of GRBs, several tight correlations have also been revealed in their afterglow phase, particularly those associated with plateau features. \cite{Dainotti..2008} first discovered a correlation between the X-ray plateau luminosity ($L_X$) and its duration ($T_a$), which is also known as the Dainotti relation (see also \citep{Dainotti..2013}). Building on this, \cite{Dainotti..2016} introduced a three-parameter correlation that includes the prompt peak luminosity ($L_{\rm p}$) in addition to $L_X$ and $T_a$. They demonstrated that this extended correlation (the Dainotti fundamental plane correlation) becomes even tighter when applied to class-specific GRB samples \citep{Dainotti..2016,Dainotti..2017ApJ}. To further reduce observational biases, \cite{Dainotti..2020ApJ} applied the \cite{Efron..1992} method to remove redshift evolution effects from each parameter. This analysis yielded a refined relation: $L_{X} \propto T_{a}^{-0.86\pm0.13} L_{\rm{p}}^{0.56\pm0.12}$, with an intrinsic scatter of about 0.22. Notably, this correlation has also proven useful for distinguishing between different GRB subclasses \citep{Lenart..2025}. Similar correlations have been noticed in GRBs exhibiting optical plateaus, in both two-parameter and three-parameter forms \citep{Dainotti..2020ApJL,Dainotti..2022ApJS,Dainotti..2022MNRAS} as well as in those exhibiting radio plateaus \citep{Levine..2022}. 
	
These tight correlations offer valuable opportunities for cosmological applications, further strengthen GRBs as cosmological probes
\citep{Dai..2004,Wang..2016,Dainotti..2022,Liang..2022,Wang..2022,Jia..2022,Dainotti..2023,Tian..2023,Hu..2023,LiJL..2024}.
In particular, these GRB correlations could provide an independent
measurement of the Hubble constant. Furthermore, GRBs at high
redshifts could potentially fill the redshift gap between SNe Ia
and the cosmic microwave background \citep{Xu..2021}.

However, a significant limitation is that the redshift of many GRBs, especially those at high redshift, is still unknown due to the difficulty in observing their host galaxies. To compensate for this weakness, GRB correlations have been used to estimate pseudo-redshifts \citep{Yonetoku..2004,Dainotti..2011,Deng..2023}, though the accuracy of this approach remains limited. An alternative solution is offered by machine learning \citep{Morgan..2012,Ukwatta..2016}. Recently, \cite{Dainotti..2024ApJL} employed a supervised statistical learning model that incorporates features from both the prompt and afterglow phases. Their predicted redshift shows strong correlation with the observed redshift, validating its reliability (see also \cite{Dainotti..2024ApJS,Dainotti..2025ApJS}). Based on this framework, \cite{Narendra..2024} released a publicly available Web-App for GRB redshift estimation, making the approach more convenient. 

The number of XRFs in our sample is still very limited, primarily
due to the lack of well-constrained spectral parameters of many
GRBs. The peak energy of XRFs is typically very close to the
low-energy threshold of most GRB detectors. Luckily, the
\emph{Einstein Probe} (\emph{EP}; \cite{Yuan..2022}) launched in
$2024$ operates in the soft X-ray band, offering new opportunities
for the study of XRFs. The Wide-field X-ray Telescope (WXT,
$0.5-4$ keV) onboard \emph{EP} has a sensitivity of $(2-3) \times
10^{-11}$ erg cm$^{-2}$ s$^{-1}$ at $1$ ks exposure, which is more
than an order of magnitude better than that of current orbiting
instruments \citep{Gao..2024,Zhang..2025}. A number of GRBs have
already been recorded by \emph{EP}/WXT
\citep{Yin..2024,Zhou..2024a,Zhou..2024b,Zhang..2025}. Another
prominent satellite is the \emph{Space-based multi-band
astronomical Variable Objects Monitor} (\emph{SVOM}; \cite{Wei..2016})
satellite which was launched in $2024$. The ECLAIRs telescope and
Gamma-Ray Monitor (GRM) onboard \emph{SVOM} are designed to
observe GRB prompt emission in the 4 -- 150 keV and 15 -- 5000 keV
energy bands, respectively \citep{Wang..2024}. With the help of
\emph{EP} and \emph{SVOM}, we would expect more XRFs with
well-constrained spectral parameters. The enlarged sample will
allow for further testing of the robustness of the correlations
revealed in this study.

\acknowledgments
% We would like to thank the anonymous referee for helpful suggestions that lead to an overall improvement of this study.
% We also thank xxx for stimulating discussion.
This study is supported
by National Key R\&D Program of China (2021YFA0718500),
by National SKA Program of China No. 2020SKA0120300,
and by the National Natural Science Foundation of China (Grant Nos. 12233002, 12273113).
YFH also acknowledges the support from the Xinjiang Tianchi Program.
JJG acknowledges support from the Youth Innovation Promotion Association (2023331).
This work made use of data supplied by the UK \emph{Swift} Science Data Centre at the
University of Leicester and the \emph{Swift} satellite.

\nocite{*}
\bibliographystyle{aasjournal}
\bibliography{bibtex}

\begin{table}[h!]
    \renewcommand{\thetable}{\arabic{table}}
    \centering
    \tabcolsep=2pt
    \renewcommand\arraystretch{1.4}
    \caption{ $303$ GRBs Detected by \emph{Swift}. } \label{tab:1}
    \begin{tabular}{lcccccccc}
        %           \tablewidth{0pt}
        \hline
        \hline
        Name & SR$^{a}$ & Redshift & $T_{90}$ & $S(25-50~\mathrm{keV})^{b}$ & $S(50-100~\mathrm{keV})^{b}$ & $S(15-150~\mathrm{keV})^{b}$ & $F(15-150~\mathrm{keV})^{c}$ & Type$^{d}$ \\
        &           &           & (s)   & ($10^{-7}$ erg cm$^{-2}$) & ($10^{-7}$ erg cm$^{-2}$) & ($10^{-7}$ erg cm$^{-2}$) & ($10^{-7}$ erg cm$^{-2}$ s$^{-1}$) &  \\
        \hline
        XRF 201015A &   2.03 $\pm$ 1.71     &   0.426   &   9.78 $\pm$ 3.47     &   0.66 $\pm$ 0.24     &   0.32 $\pm$ 0.24     &   1.97 $\pm$ 0.63     &   0.78 $\pm$ 0.22     &   Long \\
        XRF 190829A &   1.47 $\pm$ 0.35     &   0.079   &   56.90 $\pm$ 47.00   &   21.20 $\pm$ 2.27    &   14.40 $\pm$ 3.03    &   63.70 $\pm$ 7.23    &   7.42 $\pm$ 1.17     &   Long \\
        XRF 060923B &   1.38 $\pm$ 0.36     &   1.51    &   8.95 $\pm$ 1.30     &   1.62 $\pm$ 0.19     &   1.17 $\pm$ 0.27     &   4.90 $\pm$ 0.63     &   0.74 $\pm$ 0.19     &   Long \\
        XRF 210210A &   1.38 $\pm$ 0.17     &   0.715   &   6.60 $\pm$ 0.59     &   3.49 $\pm$ 0.20     &   2.53 $\pm$ 0.28     &   10.60 $\pm$ 0.66    &   3.97 $\pm$ 0.30     &   Long \\
        XRF 081007  &   1.34 $\pm$ 0.29     &   0.529   &   9.73 $\pm$ 4.87     &   2.50 $\pm$ 0.24     &   1.86 $\pm$ 0.35     &   7.61 $\pm$ 0.82     &   1.63 $\pm$ 0.24     &   Long \\
        XRR 091018  &   1.24 $\pm$ 0.07     &   0.971   &   4.37 $\pm$ 0.60     &   4.91 $\pm$ 0.14     &   3.97 $\pm$ 0.19     &   15.20 $\pm$ 0.44    &   5.91 $\pm$ 0.20     &   Long \\
        XRR 120815A &   1.23 $\pm$ 0.32     &   2.359   &   7.23 $\pm$ 2.52     &   1.59 $\pm$ 0.20     &   1.29 $\pm$ 0.29     &   4.92 $\pm$ 0.69     &   1.41 $\pm$ 0.23     &   Long \\
        XRR 090726  &   1.22 $\pm$ 0.26     &   2.71    &   56.68 $\pm$ 12.17   &   2.53 $\pm$ 0.24     &   2.08 $\pm$ 0.39     &   7.86 $\pm$ 0.89     &   0.53 $\pm$ 0.13     &   Long \\
        XRR 100316B &   1.21 $\pm$ 0.24     &   1.18    &   3.84 $\pm$ 0.43     &   0.64 $\pm$ 0.06     &   0.53 $\pm$ 0.09     &   1.98 $\pm$ 0.21     &   0.78 $\pm$ 0.11     &   Long \\
        XRR 121211A &   1.20 $\pm$ 0.34     &   1.023   &   182.70 $\pm$ 38.73  &   4.06 $\pm$ 0.50     &   3.38 $\pm$ 0.85     &   12.70 $\pm$ 1.88    &   0.98 $\pm$ 0.28     &   Long \\
        \hline
    \end{tabular}
    \begin{flushleft}
        \textbf{Notes.}
        \tablenotetext{a}{SR is the fluence ratio of $S(25-50~\mathrm{keV})$/$S(50-100~\mathrm{keV})$. }
        \tablenotetext{b}{Energy fluence in the band of $25-50$ keV ($S(25-50~\mathrm{keV})$),
               $50-100$ keV ($S(50-100~\mathrm{keV})$), and $15-150$ keV ($S(15-150~\mathrm{keV})$).
               Data are taken from the BAT3 catalog \citep{Lien..2016}. }
        \tablenotetext{c}{Peak energy flux in $15-150$ keV band. Data are taken from the BAT3 catalog \citep{Lien..2016}. }
        \tablenotetext{d}{Long or short GRBs as judged from the duration. }
        (This table is available in its entirety in machine-readable form.)
    \end{flushleft}
\end{table}

\begin{table}[h!]
    \renewcommand{\thetable}{\arabic{table}}
    \centering
    \tabcolsep=18pt
    \renewcommand\arraystretch{1.3}
    \caption{Spectral Parameters of the GRBs. }
    \label{tab:2}
    \begin{tabular}{lccccc}
        %           \tablewidth{0pt}
        \hline
        \hline
        Name & $E_{\mathrm{p, obs}}/\mathrm{keV}$ & $\alpha^{a}$ & $\beta^{b}$ & Spectra model$^{c}$ & Ref.$^{d}$ \\
        \hline
        XRF 201015A &   14 $\pm$ 6          &   -1.00 (fixed)       &   -2.40 $\pm$ 0.21    &   BAND    &   4 \\
        XRF 190829A &   11 $\pm$ 1          &   -0.92 $\pm$ 0.62    &   -2.51 $\pm$ 0.01    &   BAND    &   5 \\
        XRF 060923B &   47 $\pm$ 5.5        &   -0.11 $\pm$ 0.79    &                       &   CPL     &   2 \\
        XRF 210210A &   16.6 $\pm$ 9        &   -1.68 $\pm$ 0.24    &                       &   CPL     &   6 \\
        XRF 081007  &   40 $\pm$ 10         &   -1.40 $\pm$ 0.40    &                       &   CPL     &   7 \\
        XRR 091018  &   28 $\pm$ 13         &   -1.53 $\pm$ 0.49    &                       &   CPL     &   8 \\
        XRR 120815A &   31 $\pm$ 6.5        &   -0.87 $\pm$ 0.72    &                       &   CPL     &   2 \\
        XRR 090726  &   29 $\pm$ 6          &   -1.03 $\pm$ 0.59    &                       &   CPL     &   2 \\
        XRR 100316B &   31 $\pm$ 14         &   -1.51 $\pm$ 0.46    &                       &   CPL     &   2 \\
        XRR 121211A &   98.039 $\pm$ 10.339 &   -0.22 $\pm$ 0.31    &                       &   CPL     &   1 \\
        \hline
    \end{tabular}
    \begin{flushleft}
        \textbf{Notes.}
        \tablenotetext{a}{Low energy photon index.}
        \tablenotetext{b}{High energy photon index.}
        \tablenotetext{c}{The best-fit spectra model. BAND means the spectrum is best
                 described by a Band function \citep{Band..1993}, while CPL means the
                 spectrum is best fitted by a cut-off power-law function. }
        \tablenotetext{d}{References of the spectral parameters.
            (1) \emph{Fermi} online catalog, \url{https://heasarc.gsfc.nasa.gov/W3Browse/fermi/fermigbrst.html};
            (2) Konus-\emph{Wind} online catalog, \url{https://www.ioffe.ru/LEA/catalogs.html};
            (3) The BAT3 catalog \url{https://swift.gsfc.nasa.gov/results/batgrbcat/index\_tables.html};
            (4) \cite{Fletcher--et--al.----2020GCN.28663....1F};
            (5) \cite{Lesage--et--al.----2019GCN.25575....1L};
            (6) \cite{Frederiks--et--al.----2021GCN.29517....1F};
            (7) \cite{Bissaldi--et--al.----2008GCN..8369....1B};
            (8) \cite{Golenetskii--et--al.----2009GCN.10045....1G};
            (9) \cite{Fletcher--et--al.----2024GCN.35693....1F};
            (10) \cite{Jenke----2013GCN.15261....1J};
            (11) \cite{Stanbro--et--al.----2017GCN.22277....1S};
            (12) \cite{Veres--et--al.----2018GCN.23053....1V};
            (13) \cite{Poolakkil--et--al.----2022GCN.32089....1P};
            (14) \cite{Bissaldi--et--al.----2019GCN.26000....1B};
            (15) \cite{Veres--.---.--Fermi-GBM--Team----2022GCN.31487....1V};
            (16) \cite{Poolakkil--et--al.----2019GCN.25130....1P};
            (17) \cite{Lesage--et--al.----2020GCN.28748....1L};
            (18) \cite{Smith--et--al.----2024GCN.37478....1S};
            (19) \cite{Veres--et--al.----2024GCN.35755....1V};
            (20) \cite{Svinkin--et--al.----2024GCN.36584....1S};
            (21) \cite{von--Kienlin----2008GCN..8505....1V};
            (22) \cite{Younes--.---.--Bhat----2013GCN.14219....1Y};
            (23) \cite{Tsvetkova--et--al.----2019GCN.23637....1T};
            (24) \cite{Hamburg--et--al.----2020GCN.29140....1H};
            (25) \cite{Bissaldi--et--al.----2008GCN..8263....1B};
            (26) \cite{Wood--.---.--Fermi--GBM--Team----2021GCN.30490....1W};
            (27) \cite{Malacaria--et--al.----2021GCN.29246....1M};
            (28) \cite{Veres--et--al.----2021GCN.30779....1V};
            (29) \cite{von--Kienlin----2018GCN.23320....1V};
            (30) \cite{Myers--.---.--Fermi--Gamma-ray--Burst--Monitor--Team----2024GCN.36120....1M};
            (31) \cite{Hui----2019GCN.24002....1H};
            (32) \cite{Malacaria--et--al.----2020GCN.29073....1M};
            (33) \cite{Hamburg--et--al.----2019GCN.23707....1H};
            (34) \cite{Veres--et--al.----2023GCN.34501....1V};
            (35) \cite{Frederiks--et--al.----2022GCN.32472....1F};
            (36) \cite{Veres--et--al.----2021GCN.30233....1V};
            (37) \cite{Poolakkil--et--al.----2021GCN.30279....1P};
            (38) \cite{Veres--.---.--Fermi-GBM--Team----2020GCN.29110....1V};
            (39) \cite{de--Barra--et--al.----2023GCN.35131....1D};
            (40) \cite{Frederiks--et--al.----2021GCN.30694....1F};
            (41) \cite{Lesage--et--al.----2022GCN.31360....1L};
            (42) \cite{Sharma--et--al.----2024GCN.37301....1S};
            (43) \cite{Veres----2018GCN.23352....1V};
            (44) \cite{Frederiks--et--al.----2019GCN.26576....1F};
            (45) \cite{Frederiks--et--al.----2018GCN.22546....1F};
            (46) \cite{Lesage--et--al.----2022GCN.33112....1L};
            (47) \cite{Ridnaia--et--al.----2024GCN.37139....1R};
            (48) \cite{Malacaria--et--al.----2021GCN.30199....1M};
            (49) \cite{Svinkin--et--al.----2019GCN.25974....1S};
            (50) \cite{Frederiks--et--al.----2023GCN.35359....1F};
            (51) \cite{Bissaldi--et--al.----2023GCN.35369....1B};
            (52) \cite{Lesage--et--al.----2020GCN.28326....1L};
            (53) \cite{Lesage--et--al.----2021GCN.30573....1L}; }
        (This table is available in its entirety in machine-readable form.)
    \end{flushleft}
\end{table}

\begin{figure}[htbp]
    \centering
    \includegraphics[width=0.8\linewidth]{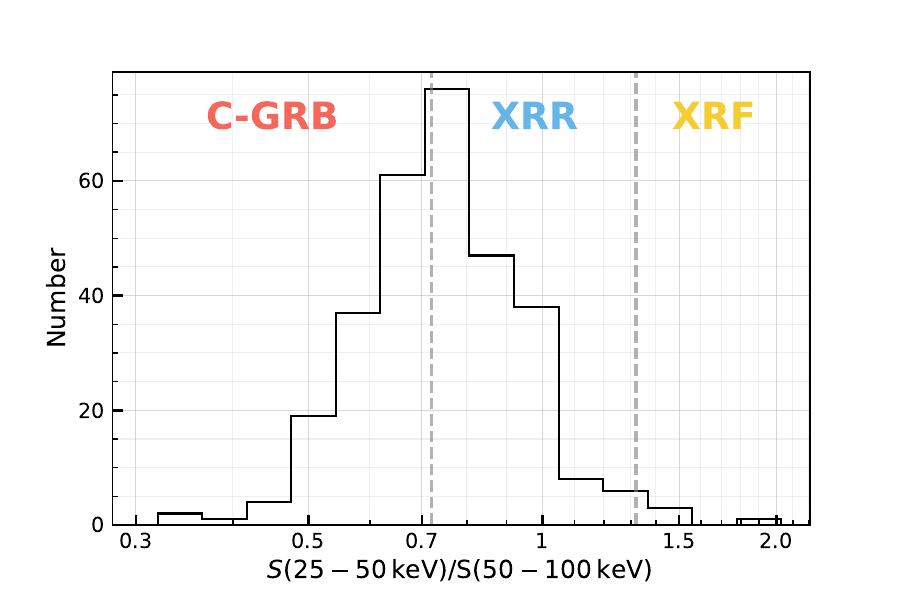}
    \caption{The distribution of the fluence ratio $S(25-50~\mathrm{keV}) / S(50-100~\mathrm{keV})$
       for the whole sample. The vertical dashed lines correspond to the borders between
       C-GRBs and XRRs, and between XRRs and XRFs. } \label{fig:1}
\end{figure}

\begin{table}[h!]
    \renewcommand{\thetable}{\arabic{table}}
    \centering
    \tabcolsep=15pt
    \renewcommand\arraystretch{1.2}
    \caption{Some Key Parameters of the GRBs.  }
    \label{tab:3}
    \begin{tabular}{lccccc}
        %           \tablewidth{0pt}
        \hline
        \hline
        Name    &   $\log(\mathrm{SR})^{a}$ &   $\log(T_{90, \mathrm{rest}}/\mathrm{s})^{b}$    &   $\log(E_{\mathrm{iso}}/\mathrm{erg})^{c}$   &   $\log(L_{\mathrm{p}}/(\mathrm{erg/s}))^{c}$ &   $\log(E_{\mathrm{p}}/\mathrm{keV})^{d}$ \\
        \hline
        XRF 201015A &   0.31 $\pm$ 0.36     &   0.84 $\pm$ 0.15     &   50.39 $\pm$ 0.14    &   50.14 $\pm$ 0.12    &   1.30 $\pm$ 0.19  \\
        XRF 190829A &   0.17 $\pm$ 0.10     &   1.72 $\pm$ 0.36     &   50.55 $\pm$ 0.05    &   49.65 $\pm$ 0.07    &   1.07 $\pm$ 0.04  \\
        XRF 060923B &   0.14 $\pm$ 0.11     &   0.55 $\pm$ 0.06     &   51.55 $\pm$ 0.06    &   51.13 $\pm$ 0.11    &   2.07 $\pm$ 0.05  \\
        XRF 210210A &   0.14 $\pm$ 0.05     &   0.59 $\pm$ 0.04     &   51.58 $\pm$ 0.03    &   51.39 $\pm$ 0.03    &   1.45 $\pm$ 0.24  \\
        XRF 081007  &   0.13 $\pm$ 0.09     &   0.80 $\pm$ 0.22     &   51.01 $\pm$ 0.05    &   50.52 $\pm$ 0.06    &   1.79 $\pm$ 0.11  \\
        XRR 091018  &   0.09 $\pm$ 0.02     &   0.35 $\pm$ 0.06     &   51.91 $\pm$ 0.01    &   51.79 $\pm$ 0.01    &   1.74 $\pm$ 0.20  \\
        XRR 120815A &   0.09 $\pm$ 0.11     &   0.33 $\pm$ 0.15     &   52.02 $\pm$ 0.06    &   52.00 $\pm$ 0.07    &   2.02 $\pm$ 0.09  \\
        XRR 090726  &   0.09 $\pm$ 0.09     &   1.18 $\pm$ 0.09     &   52.36 $\pm$ 0.05    &   51.76 $\pm$ 0.10    &   2.03 $\pm$ 0.09  \\
        XRR 100316B &   0.08 $\pm$ 0.09     &   0.25 $\pm$ 0.05     &   51.17 $\pm$ 0.05    &   51.11 $\pm$ 0.06    &   1.83 $\pm$ 0.20  \\
        XRR 121211A &   0.08 $\pm$ 0.12     &   1.96 $\pm$ 0.09     &   51.69 $\pm$ 0.06    &   50.88 $\pm$ 0.13    &   2.30 $\pm$ 0.05 \\
        \hline
    \end{tabular}
    \begin{flushleft}
        \textbf{Notes.}
        \tablenotetext{a}{The logrithm value of the fluence ratio of $S(25-50~\mathrm{keV})$/$S(50-100~\mathrm{keV})$. }
        \tablenotetext{b}{The rest frame burst duration, calculated with $T_{90, \mathrm{rest}}=T_{90}/(1+z)$. }
        \tablenotetext{c}{Converted to the rest frame energy band of $1-10^{4}$ keV. }
        \tablenotetext{d}{The rest frame peak energy, calculated with $E_{\mathrm{p}}=E_{\mathrm{p, obs}}\times(1+z)$. }
        (This table is available in its entirety in machine-readable form.)
    \end{flushleft}
\end{table}

\begin{table}[h!]
    \renewcommand{\thetable}{\arabic{table}}
    \centering
    \tabcolsep=10pt
    \renewcommand\arraystretch{1.2}
    \caption{K-S Test Results. } \label{tab:4}
    \begin{tabular}{lcccccccc}
        %           \tablewidth{0pt}
        \hline
        \hline
        Sub-samples &   $E_{\mathrm{p, obs}}$  &   $\alpha$    &   $\beta$ &   $\log(\mathrm{SR})$ &   $\log(T_{90, \mathrm{rest}})$    &   $\log(E_{\mathrm{iso}})$   &   $\log(L_{\mathrm{p}})$ &   $\log(E_{\mathrm{p}})$ \\
        \hline
        LGRB \& SGRB    &   0.00032     &   3.10E-05    &           &   0.0038      &   1.60E-10    &   1.70E-11    &   0.00018     &   0.26 \\
        XRR \& C-GRB    &   2.80E-22    &   3.70E-13    &   0.8     &   7.50E-88    &   0.1         &   4.70E-06    &   2.50E-07    &   1.10E-23 \\
        XRF \& XRR      &   0.0002      &   0.65        &   0.34    &   1.70E-09    &   0.079       &   9.00E-05    &   0.0024      &   0.00011 \\
        XRF \& C-GRB    &   5.60E-09    &   0.36        &   0.46    &   5.60E-09    &   0.094       &   2.40E-05    &   0.00024     &   1.20E-07 \\
        \hline
    \end{tabular}
    \begin{flushleft}
        \textbf{Notes.} Pairvise comparison of various sub-samples.
         K-S test probability ($P$) for different parameters are listed.
    \end{flushleft}
\end{table}

\begin{table}[h!]
    \renewcommand{\thetable}{\arabic{table}}
    \centering
    \tabcolsep=4pt
    \renewcommand\arraystretch{1.4}
    \caption{The best-fit results of various correlations. }
    \label{tab:5}
    \begin{tabular}{lccccccc}
        %           \tablewidth{0pt}
        \hline
        \hline
        \multicolumn{8}{c}{All GRBs} \\
        \hline
        Correlation form    &   $a$ &   $b$ &   $c$ &   $\sigma_{\mathrm{in}}^{a}$  &   $\chi^{2}/\mathrm{d.o.f.}^{b}$  &   $\rho^{c}$  &   $p^{c}$ \\
        \hline
        $\log E_{\mathrm{p}} = a + b \log L_{\mathrm{p}}$   &   -10.977 $\pm$ 1.056 &   0.261 $\pm$ 0.02    &   &   0.306 $\pm$ 0.014   &   0.108   &   0.61    &   7.80E-33 \\
        $\log E_{\mathrm{p}} = a + b \log E_{\mathrm{iso}}$ &   -10.639 $\pm$ 1.073 &   0.252 $\pm$ 0.02    &   &   0.312 $\pm$ 0.014   &   0.113   &   0.61    &   9.40E-33 \\
        $\log L_{\mathrm{p}} = a + b \log E_{\mathrm{iso}}$ &   7.085 $\pm$ 1.49    &   0.854 $\pm$ 0.028   &   &   0.456 $\pm$ 0.019   &   0.224   &   0.85    &   7.60E-84 \\
        $\log E_{\mathrm{p}} = a + b \log \mathrm{SR}$  &   2.25 $\pm$ 0.026    &   -2.914 $\pm$ 0.158  &   &   0.21 $\pm$ 0.012    &   0.075   &   -0.74   &   1.30E-53 \\
        $\log E_{\mathrm{iso}} = a + b \log \mathrm{SR}$    &   52.325 $\pm$ 0.082  &   -2.881 $\pm$ 0.529  &   &   0.872 $\pm$ 0.037   &   0.813   &   -0.36   &   6.10E-11 \\
        $\log L_{\mathrm{p}} = a + b \log \mathrm{SR}$  &   51.689 $\pm$ 0.081  &   -2.893 $\pm$ 0.515  &   &   0.861 $\pm$ 0.037   &   0.796   &   -0.36   &   7.70E-11 \\
        $\log E_{\mathrm{iso}} = a + b \log L_{\mathrm{p}} + c \log T_{90, \mathrm{rest}}$  &   6.015 $\pm$ 0.836   &   0.883 $\pm$ 0.016   &   0.582 $\pm$ 0.021   &   0.235 $\pm$ 0.011   &   0.065   &   0.95    &   6.20E-153 \\
        $\log E_{\mathrm{p}} = a + b \log L_{\mathrm{p}} + c \log T_{90, \mathrm{rest}}$    &   -11.047 $\pm$ 1.058 &   0.262 $\pm$ 0.02    &   0.041 $\pm$ 0.028   &   0.306 $\pm$ 0.014   &   0.108   &   0.62    &   6.70E-34 \\
        $\log E_{\mathrm{p}} = a + b \log E_{\mathrm{iso}} + c \log T_{90, \mathrm{rest}}$  &   -12.486 $\pm$ 1.126 &   0.29 $\pm$ 0.022    &   -0.131 $\pm$ 0.03   &   0.301 $\pm$ 0.014   &   0.107   &   0.63    &   2.50E-35 \\
        $\log E_{\mathrm{p}} = a + b \log E_{\mathrm{iso}} + c \log L_{\mathrm{p}}$ &   -11.773 $\pm$ 1.069 &   0.108 $\pm$ 0.039   &   0.168 $\pm$ 0.04    &   0.303 $\pm$ 0.014   &   0.106   &   0.64    &   8.40E-36 \\
        $\log E_{\mathrm{p}} = a + b \log E_{\mathrm{iso}} + c \log \mathrm{SR}$    &   -7.057 $\pm$ 0.772  &   0.177 $\pm$ 0.015   &   -2.398 $\pm$ 0.131  &   0.169 $\pm$ 0.01    &   0.052   &   0.84    &   2.70E-80 \\
        $\log E_{\mathrm{p}} = a + b \log L_{\mathrm{p}} + c \log \mathrm{SR}$  &   -6.771 $\pm$ 0.792  &   0.174 $\pm$ 0.015   &   -2.336 $\pm$ 0.136  &   0.175 $\pm$ 0.01    &   0.051   &   0.84    &   1.50E-80 \\
        \hline
        \hline
        \multicolumn{8}{c}{LGRBs only} \\
        \hline
        Correlation form    &   $a$ &   $b$ &   $c$ &   $\sigma_{\mathrm{in}}^{a}$  &   $\chi^{2}/\mathrm{d.o.f.}^{b}$  &   $\rho^{c}$  &   $p^{c}$ \\
        \hline
        $\log E_{\mathrm{p}} = a + b \log L_{\mathrm{p}}$   &   -23.336 $\pm$ 0.798 &   0.496 $\pm$ 0.015   &   &   0.278 $\pm$ 0.014   &   0.143   &   0.67    &   2.10E-37 \\
        $\log E_{\mathrm{p}} = a + b \log E_{\mathrm{iso}}$ &   -23.477 $\pm$ 0.823 &   0.493 $\pm$ 0.016   &   &   0.271 $\pm$ 0.014   &   0.111   &   0.69    &   1.70E-41 \\
        $\log L_{\mathrm{p}} = a + b \log E_{\mathrm{iso}}$ &   2.621 $\pm$ 1.709   &   0.938 $\pm$ 0.032   &   &   0.441 $\pm$ 0.019   &   0.209   &   0.84    &   2.20E-76 \\
        $\log E_{\mathrm{p}} = a + b \log \mathrm{SR}$  &   2.235 $\pm$ 0.027   &   -3.109 $\pm$ 0.166  &   &   0.195 $\pm$ 0.012   &   0.063   &   -0.76   &   2.50E-53 \\
        $\log E_{\mathrm{iso}} = a + b \log \mathrm{SR}$    &   52.323 $\pm$ 0.068  &   -4.034 $\pm$ 0.46   &   &   0.698 $\pm$ 0.033   &   0.54    &   -0.46   &   2.20E-16 \\
        $\log L_{\mathrm{p}} = a + b \log \mathrm{SR}$  &   51.669 $\pm$ 0.08   &   -3.705 $\pm$ 0.52   &   &   0.8 $\pm$ 0.037 &   0.7 &   -0.42   &   1.20E-13 \\
        $\log E_{\mathrm{iso}} = a + b \log L_{\mathrm{p}} + c \log T_{90, \mathrm{rest}}$  &   6.372 $\pm$ 0.893   &   0.876 $\pm$ 0.017   &   0.602 $\pm$ 0.026   &   0.226 $\pm$ 0.011   &   0.06    &   0.94    &   8.40E-138 \\
        $\log E_{\mathrm{p}} = a + b \log L_{\mathrm{p}} + c \log T_{90, \mathrm{rest}}$    &   -24.011 $\pm$ 0.799 &   0.506 $\pm$ 0.015   &   0.124 $\pm$ 0.032   &   0.269 $\pm$ 0.014   &   0.081   &   0.68    &   2.20E-40 \\
        $\log E_{\mathrm{p}} = a + b \log E_{\mathrm{iso}} + c \log T_{90, \mathrm{rest}}$  &   -24.071 $\pm$ 0.786 &   0.508 $\pm$ 0.015   &   -0.177 $\pm$ 0.03   &   0.251 $\pm$ 0.013   &   0.075   &   0.71    &   7.60E-44 \\
        $\log E_{\mathrm{p}} = a + b \log E_{\mathrm{iso}} + c \log L_{\mathrm{p}}$ &   -24.288 $\pm$ 0.794 &   0.28 $\pm$ 0.039    &   0.231 $\pm$ 0.039   &   0.252 $\pm$ 0.013   &   0.074   &   0.71    &   4.00E-44 \\
        $\log E_{\mathrm{p}} = a + b \log E_{\mathrm{iso}} + c \log \mathrm{SR}$    &   -8.146 $\pm$ 0.92   &   0.198 $\pm$ 0.018   &   -2.274 $\pm$ 0.153  &   0.165 $\pm$ 0.01    &   0.044   &   0.85    &   1.00E-81 \\
        $\log E_{\mathrm{p}} = a + b \log L_{\mathrm{p}} + c \log \mathrm{SR}$  &   -6.57 $\pm$ 0.841   &   0.17 $\pm$ 0.016    &   -2.391 $\pm$ 0.153  &   0.169 $\pm$ 0.01    &   0.043   &   0.86    &   5.30E-82 \\
        \hline
    \end{tabular}
    \begin{flushleft}
        \textbf{Notes.}
        \tablenotetext{a}{Intrinsic scatter of the correlation obtained from MCMC fitting. }
        \tablenotetext{b}{The reduced $\chi^2$ of the best-fit correlation. }
        \tablenotetext{c}{$\rho$ and $p$ represent the Spearman correlation coefficient and the corresponding chance probability, respectively. }
    \end{flushleft}
\end{table}

\begin{table}[h!]
	\renewcommand{\thetable}{\arabic{table}}
	\centering
	\tabcolsep=4pt
	\renewcommand\arraystretch{1.4}
	\caption{The best-fit results of various correlations after correcting redshift evolution. }
	\label{tab:6}
	\begin{tabular}{lccccccc}
		%           \tablewidth{0pt}
		\hline
		\hline
		\multicolumn{8}{c}{All GRBs} \\
		\hline
		Correlation form$^{a}$    &   $a$ &   $b$ &   $c$ &   $\sigma_{\mathrm{in}}$  &   $\chi^{2}/\mathrm{d.o.f.}$  &   $\rho$  &   $p$ \\
		\hline
		$\log E^{\prime}_{\mathrm{p}} = a + b \log L^{\prime}_{\mathrm{p}}$   & -11.66 $\pm$ 1.36  & 0.276 $\pm$ 0.027 &       & 0.306 $\pm$ 0.014 & 0.109 & 0.5   & 5.50E-21 \\
		$\log E^{\prime}_{\mathrm{p}} = a + b \log E^{\prime}_{\mathrm{iso}}$ & -9.867 $\pm$ 1.263 & 0.235 $\pm$ 0.024 &       & 0.313 $\pm$ 0.014 & 0.112 & 0.49  & 1.80E-19 \\
		$\log L^{\prime}_{\mathrm{p}} = a + b \log E^{\prime}_{\mathrm{iso}}$ & 15.125 $\pm$ 1.67  & 0.685 $\pm$ 0.032 &       & 0.425 $\pm$ 0.018 & 0.197 & 0.72  & 1.20E-49 \\
		$\log E^{\prime}_{\mathrm{p}} = a + b \log \mathrm{SR}$  & 1.924 $\pm$ 0.022 & -2.887 $\pm$ 0.134 &       & 0.17  $\pm$ 0.011 & 0.055 & -0.79 & 3.60E-65 \\
		$\log E^{\prime}_{\mathrm{iso}} = a + b \log \mathrm{SR}$    & 51.335 $\pm$ 0.066 & -2.556 $\pm$ 0.434 &       & 0.71  $\pm$ 0.031 & 0.538 & -0.37 & 2.20E-11 \\
		$\log L^{\prime}_{\mathrm{p}} = a + b \log \mathrm{SR}$  & 50.186 $\pm$ 0.058 & -2.404 $\pm$ 0.373 &       & 0.623 $\pm$ 0.027 & 0.417 & -0.41 & 1.30E-13 \\
		$\log E^{\prime}_{\mathrm{iso}} = a + b \log L^{\prime}_{\mathrm{p}} + c \log T^{\prime}_{90, \mathrm{rest}}$  & 8.089 $\pm$ 1.122 & 0.845 $\pm$ 0.022 & 0.603 $\pm$ 0.022 & 0.247 $\pm$ 0.011 & 0.071 & 0.92  & 1.90E-126 \\
		$\log E^{\prime}_{\mathrm{p}} = a + b \log L^{\prime}_{\mathrm{p}} + c \log T^{\prime}_{90, \mathrm{rest}}$    & -11.631 $\pm$ 1.35  & 0.274 $\pm$ 0.027 & 0.041 $\pm$ 0.028 & 0.305 $\pm$ 0.014 & 0.108 & 0.51  & 1.80E-21 \\
		$\log E^{\prime}_{\mathrm{p}} = a + b \log E^{\prime}_{\mathrm{iso}} + c \log T^{\prime}_{90, \mathrm{rest}}$  & -13.338 $\pm$ 1.463 & 0.307 $\pm$ 0.029 & -0.145 $\pm$ 0.033 & 0.301 $\pm$ 0.014 & 0.107 & 0.52  & 1.60E-22 \\
		$\log E^{\prime}_{\mathrm{p}} = a + b \log E^{\prime}_{\mathrm{iso}} + c \log L^{\prime}_{\mathrm{p}}$ & -12.518 $\pm$ 1.373 & 0.109 $\pm$ 0.038 & 0.182 $\pm$ 0.042 & 0.302 $\pm$ 0.014 & 0.106 & 0.52  & 9.70E-23 \\
		$\log E^{\prime}_{\mathrm{p}} = a + b \log E^{\prime}_{\mathrm{iso}} + c \log \mathrm{SR}$    & -4.637 $\pm$ 0.863 & 0.127 $\pm$ 0.017 & -2.532 $\pm$ 0.128 & 0.154 $\pm$ 0.01  & 0.048 & 0.82  & 4.10E-74 \\
		$\log E^{\prime}_{\mathrm{p}} = a + b \log L^{\prime}_{\mathrm{p}} + c \log \mathrm{SR}$  & -3.966 $\pm$ 0.986 & 0.117 $\pm$ 0.02  & -2.515 $\pm$ 0.141 & 0.164 $\pm$ 0.01  & 0.048 & 0.81  & 6.80E-73 \\
		\hline
		\hline
		\multicolumn{8}{c}{LGRBs only} \\
		\hline
		Correlation form$^{a}$    &   $a$ &   $b$ &   $c$ &   $\sigma_{\mathrm{in}}$  &   $\chi^{2}/\mathrm{d.o.f.}$  &   $\rho$  &   $p$ \\
		\hline
		$\log E^{\prime}_{\mathrm{p}} = a + b \log L^{\prime}_{\mathrm{p}}$   & -21.186 $\pm$ 1.35  & 0.459 $\pm$ 0.027 &       & 0.273 $\pm$ 0.018 & 0.197 & 0.51  & 6.70E-20 \\
		$\log E^{\prime}_{\mathrm{p}} = a + b \log E^{\prime}_{\mathrm{iso}}$ & -16.961 $\pm$ 1.123 & 0.368 $\pm$ 0.022 &       & 0.259 $\pm$ 0.018 & 0.123 & 0.58  & 2.90E-26 \\
		$\log L^{\prime}_{\mathrm{p}} = a + b \log E^{\prime}_{\mathrm{iso}}$ & 10.048 $\pm$ 1.718 & 0.782 $\pm$ 0.033 &       & 0.388 $\pm$ 0.017 & 0.166 & 0.76  & 2.30E-54 \\
		$\log E^{\prime}_{\mathrm{p}} = a + b \log \mathrm{SR}$  & 1.917 $\pm$ 0.023 & -2.965 $\pm$ 0.144 &       & 0.162 $\pm$ 0.01  & 0.044 & -0.8  & 1.00E-64 \\
		$\log E^{\prime}_{\mathrm{iso}} = a + b \log \mathrm{SR}$    & 51.329 $\pm$ 0.058 & -3.443 $\pm$ 0.404 &       & 0.603 $\pm$ 0.028 & 0.398 & -0.45 & 1.00E-15 \\
		$\log L^{\prime}_{\mathrm{p}} = a + b \log \mathrm{SR}$  & 50.168 $\pm$ 0.06  & -2.723 $\pm$ 0.405 &       & 0.616 $\pm$ 0.028 & 0.409 & -0.41 & 1.20E-12 \\
		$\log E^{\prime}_{\mathrm{iso}} = a + b \log L^{\prime}_{\mathrm{p}} + c \log T^{\prime}_{90, \mathrm{rest}}$  & 8.201 $\pm$ 1.11  & 0.842 $\pm$ 0.022 & 0.605 $\pm$ 0.027 & 0.236 $\pm$ 0.011 & 0.065 & 0.92  & 4.00E-118 \\
		$\log E^{\prime}_{\mathrm{p}} = a + b \log L^{\prime}_{\mathrm{p}} + c \log T^{\prime}_{90, \mathrm{rest}}$    & -21.305 $\pm$ 1.324 & 0.459 $\pm$ 0.026 & 0.084 $\pm$ 0.049 & 0.271 $\pm$ 0.018 & 0.054 & 0.53  & 2.90E-22 \\
		$\log E^{\prime}_{\mathrm{p}} = a + b \log E^{\prime}_{\mathrm{iso}} + c \log T^{\prime}_{90, \mathrm{rest}}$  & -22.292 $\pm$ 1.289 & 0.478 $\pm$ 0.026 & -0.241 $\pm$ 0.036 & 0.219 $\pm$ 0.016 & 0.057 & 0.56  & 7.60E-25 \\
		$\log E^{\prime}_{\mathrm{p}} = a + b \log E^{\prime}_{\mathrm{iso}} + c \log L^{\prime}_{\mathrm{p}}$ & -21.56 $\pm$ 1.293 & 0.165 $\pm$ 0.04  & 0.298 $\pm$ 0.052 & 0.225 $\pm$ 0.016 & 0.05  & 0.56  & 1.20E-24 \\
		$\log E^{\prime}_{\mathrm{p}} = a + b \log E^{\prime}_{\mathrm{iso}} + c \log \mathrm{SR}$    & -5.507 $\pm$ 0.975 & 0.144 $\pm$ 0.019 & -2.423 $\pm$ 0.143 & 0.15  $\pm$ 0.01  & 0.038 & 0.84  & 1.40E-75 \\
		$\log E^{\prime}_{\mathrm{p}} = a + b \log L^{\prime}_{\mathrm{p}} + c \log \mathrm{SR}$  & -3.559 $\pm$ 0.971 & 0.109 $\pm$ 0.019 & -2.578 $\pm$ 0.151 & 0.157 $\pm$ 0.01  & 0.039 & 0.83  & 2.80E-72 \\
		
		\hline
	\end{tabular}
	\begin{flushleft}
		\textbf{Notes.}
		\tablenotetext{a}{The de-evolved parameters are denoted with the symbol $^{\prime}$. }
%		\tablenotetext{b}{The reduced $\chi^2$ of the best-fit correlation. }
%		\tablenotetext{c}{$\rho$ and $p$ represent the Spearman correlation coefficient and the corresponding chance probability, respectively. }
	\end{flushleft}
\end{table}

%\appendix
%
%\section{  }

\end{document}